\documentclass[twocolumn,showpacs,eqsecnum,preprintnumbers,amsmath,amssymb]{revtex4}

\usepackage{epsfig,psfrag}
\usepackage{dcolumn}
\usepackage{bm}
\usepackage{graphicx}
\usepackage{color}
\usepackage{amsthm}

\begin{document}


\title{Three-body problem for ultracold atoms in quasi-one-dimensional traps}

\author{C.~Mora,$^1$ R.~Egger,$^1$ and  A.O.~Gogolin$^2$}

\affiliation{${}^1$~Institut f\"ur Theoretische Physik, 
Heinrich-Heine-Universit\"at,
 D-40225 D\"usseldorf, Germany\\
${}^{2}$ Department of Mathematics, Imperial College London, 
180 Queen's Gate, London SW7 2AZ, United Kingdom 
}

\date{\today}

\begin{abstract}
We study the three-body problem for both fermionic and
bosonic cold atom gases in a parabolic transverse trap
of lengthscale $a_\perp$.  For this quasi-one-dimensional (1D)
problem, there is a two-body 
bound state (dimer) for any sign of the 3D scattering length $a$, and
a confinement-induced scattering resonance.
The fermionic three-body problem is universal
and characterized by two atom-dimer scattering 
lengths, $a_{ad}$ and $b_{ad}$. In the tightly bound `dimer limit',
$a_\perp/a\to\infty$, we find $b_{ad}=0$, and $a_{ad}$ is linked to the 3D
atom-dimer scattering length.  In the weakly bound `BCS limit',
$a_\perp/a\to-\infty$, a connection to the Bethe Ansatz is established, which allows for exact results. 
The full crossover is obtained numerically.
The bosonic three-body problem, however, is non-universal:
$a_{ad}$ and $b_{ad}$ depend both
on $a_\perp/a$ and on a parameter $R^*$
related to the sharpness of the resonance.
Scattering solutions are qualitatively similar to fermionic ones.
We predict the existence of a single confinement-induced
three-body bound state (trimer) for bosons. 
\end{abstract}

\pacs{03.75.Ss, 05.30.Fk, 03.65.Nk}

\maketitle

\section{Introduction}

The physics of cold atoms has recently enjoyed a great amount
of attention.  A particularly interesting phenomenon in that
context has been the experimental observation of dimer (molecule)
formation in ultracold binary Fermi gases \cite{molecule1},
where a Feshbach resonance is exploited.  This has allowed
experimental access to the full crossover
 from a Bose-Einstein condensate (BEC)
to a BCS-type superfluid by simply tuning a magnetic field 
 \cite{regal04,grimm04,ketterle04,bourdel,chin}.  
Because of the Feshbach-resonant behavior, the 
3D scattering length $a$ describing the $s$-wave
interaction strength among different fermions can be tuned
almost at will. For $a>0$, one has a two-body bound state (`dimer') that
eventually can be Bose-condensed, while for $a<0$, Cooper pairs are formed.
The corresponding BEC-BCS crossover
theory has been worked out on a
mean-field (plus fluctuations) level \cite{randeria,timmermanns,ohashi,duine}, 
and is widely believed to account for the basic experimental observations
\cite{perali}.  The Feshbach resonance has also interesting implications
for bosonic systems, e.g., the existence of a similar crossover from  
atomic BEC to molecular BEC 
\cite{holland,donley,herbig}, 
or Bose-enhanced `quantum superchemistry' \cite{hope,moore}. 

A related but different quasi-1D problem arises
for either a two-species ($\uparrow,\downarrow$)
Fermi gas or a single-species Bose gas.
When such a cold atom gas is confined to a sufficiently
tight harmonic transverse trap, it enters a 1D regime. On the two-body level,
there is always a bound state, even for $a<0$, and one has a 
confinement-induced resonance (CIR) in the 
1D atom-atom scattering length $a_{aa}$
\cite{olshanii98,bergeman03}.  This `shape' (or `geometric') resonance
is similar to a Feshbach resonance \cite{hammer}.  Instead of a magnetic
field, the ratio $a/a_\perp$ between the 3D scattering length 
and  the transverse confinement lengthscale $a_\perp$,
see Eq.~(\ref{aperp}) below,
is now used to sweep through the resonance. 
Albeit there are strong quantum fluctuations in 1D, preventing 
both BEC and a true BCS superfluid in the thermodynamic limit,
the presence of the CIR leads to a rather similar
scenario as for the standard (3D) BEC-BCS crossover 
\cite{tokatly,fuchs,blume}. Moreover,
this 1D analogue of the 3D BEC-BCS crossover and its
bosonic complement appear to be experimentally feasible.
Recent progress towards the realization of 1D traps has been
tremendous \cite{gorlitz,esslinger,richard,paredes,weiss}
and could lead to the observation of interesting
aspects of 1D many-body physics as outlined below. 
Moreover, on the theoretical
side, powerful many-body techniques are available in 1D systems,
 e.g., bosonization \cite{book}, or the Bethe Ansatz \cite{gaudin,sutherland}.
Such methods often allow for exact statements.  
Below we analytically solve the three-body problem  
for ultracold fermions or bosons that are confined to quasi-1D by a
transverse trap potential.  We show that, for low energies,
atom-dimer scattering solutions can be completely described
in terms of two scattering lengths $a_{ad}$ and $b_{ad}$,
where $a_{ad}$ can be linked to the standard 3D atom-dimer
scattering length.  The non-standard scattering length $b_{ad}$ must 
be introduced to describe  1D scattering processes in general.  
While the (fermionic and bosonic) three-body problem in 3D
systems has a rather long history, reviewed in Refs.~\cite{hammer,nielsen},
the quasi-1D situation has not received much attention so far.

In what follows, we assume two different hyperfine states of 
a fermion species of mass $m_0$, or a single boson species,
to be trapped in the parabolic transverse confinement potential 
\begin{equation}\label{confinment}
U_c({\bf r}) = \frac12 m_0\omega_\perp^2 ( x^2+y^2),
\end{equation}
 with associated confinement lengthscale 
\begin{equation}\label{aperp}
a_\perp = \sqrt{ 2\hbar/m_0 \omega_\perp } .
\end{equation}
For a non-parabolic potential, the treatment is more 
involved \cite{peano}, and new effects appear, e.g., additional resonances.
Under the standard pseudopotential 
approximation \cite{huang}, for the low-energy limit
of interest here, the 3D interaction can be written as  
\begin{equation}\label{pseudopot}
V({\bf r})=\frac{4\pi\hbar^2 a}{m_0} \delta({\bf r}) 
\frac{\partial}{\partial r} (r \cdot),
\end{equation}
which assumes that the interaction range is the shortest lengthscale
of relevance in the problem.  
On the two-body level, interactions between atoms are thereby described
in terms of the 3D scattering length $a$.
Normally, at low energy scales only $s$-wave interactions matter.
Identical fermions then do not interact because
of the Pauli principle, but interactions among different ones
will be present and are given by Eq.~(\ref{pseudopot}).
The free (unconfined) problem is well-known to possess 
exactly one bound state (`dimer') 
on the two-body level for $a>0$ \cite{landau}.
Once the confinement (\ref{confinment}) is present, 
however, 
there is {\sl always} a two-body bound state with dimensionless binding energy 
\begin{equation}\label{omegab}
\Omega_B =  (\hbar\omega_\perp-E_B)/2\hbar\omega_\perp = (\kappa_B a_\perp/2)^2,
\end{equation}
where $a_B=1/\kappa_B$ is the (longitudinal) size of the dimer. 
This quantity is determined by the condition \cite{olshanii98,bergeman03}
\begin{equation}\label{bs}
\zeta(1/2,\Omega_B) + a_\perp/a =0.
\end{equation}
For our purposes, the Hurvitz zeta function can be defined via its integral
representation (see appendix \ref{appen}),
\begin{equation}\label{zeta}
\zeta(1/2,\Omega) = \int_0^\infty \frac{dt}{\sqrt{\pi t}}
\left( \frac{e^{-\Omega t}}{1-e^{-t}} - \frac{1}{t} \right).
\end{equation} 
Since $\zeta(1/2,\Omega)$ is monotonic in $\Omega$, 
there is precisely one bound state for any given $a_\perp/a$.
This bound-state wavefunction can then be expressed in terms of
the single-particle Green's function $G_E ({\bf r},{\bf r'})$ for the 
cylindrical harmonic oscillator with reduced mass $m_0/2$,
see Eqs.~\eqref{green1} and \eqref{tokat} below.
For $a_\perp/a\to -\infty$, the `BCS limit' is reached,
 where $\Omega_B\simeq (a/a_\perp)^2 \ll 1$. (For simplicity, we 
shall use the phrase `BCS limit' also for bosons, even though
there is no Cooper pairing in that case.)
The dimer is then very elongated, with size 
$a_B \approx a_\perp^2/|a| \gg a_\perp$ in the axial direction,
and confined to the transverse ground state.
The dimer in this limit can be effectively described by
a $1$D contact interaction obtained by projecting the pseudopotential
\eqref{pseudopot} to the lowest transverse state.
In the tightly bound `dimer limit', $a_\perp/a\to +\infty$,  
the dimer becomes spherically symmetric with the 
size $a_B\approx a \ll a_\perp$. Here, one recovers the pseudopotential
bound state of the unconfined problem (for $a>0$)  with the large 
reduced binding energy $\Omega_B\simeq (a_\perp/2a)^2\gg 1$.

The analogue of the Feshbach resonance in a 1D confined
 system is then realized by the 
CIR.  Solving the two-body scattering problem with just one open channel,
the 1D scattering length between two atoms can be extracted
and is found to be  \cite{olshanii98,bergeman03}
\begin{equation}\label{a1d}
a_{aa}=-\frac{a_\perp}{2} \left[\frac{a_\perp}{a}-{\cal C}\right],\quad
{\cal C}=-\zeta(1/2)\simeq 1.4603.
\end{equation}
At low energies, this implies that one can use
the  1D atom-atom interaction potential 
\begin{equation}\label{vaa}
V_{aa}(z,z')=g_{aa} \delta(z-z'),\quad g_{aa}=-\frac{2\hbar^2}{m_0 a_{aa}}.
\end{equation}
The CIR, where $g_{aa}\to \pm\infty$,
then occurs for $a_{aa}=0$, corresponding to $\Omega_B=1$,
and can be reached by tuning $a_\perp$ or $a$.  The physical 
picture behind the emergence of the resonance has been elucidated
in Ref.~\cite{bergeman03} and shown to be similar to a Feshbach
resonance. Atoms in the lowest transverse level (open channel)
are coupled resonantly to a bound state in the two-body
sector restricted to excited channels (closed channel).
This picture has also recently been extended to non-parabolic 
potentials \cite{peano}.
Although up to now, no clear experimental evidence for CIR behavior
has been published, there are several possibilities to observe
it using standard techniques, e.g., via the momentum distribution 
or Bragg spectroscopy.

The above considerations suggest that it is mandatory to 
analyze the consequences of this CIR on the many-body 
scattering properties \cite{tokatly,fuchs}. 
As a first step towards a full understanding of this problem, 
we discuss the analytic solution of the 
corresponding three-body problem, which here becomes possible
due to the short-rangedness of the atom-atom interaction.
A natural question then concerns the scattering properties of
the atom-dimer system, for instance, the scattering length
$a_{ad}$.  For fermions, this problem has been solved already a long
time ago by Skorniakov and Ter-Martirosian (STM) \cite{skorniakov},
who found $a_{ad}\approx 1.2 a$. (This result was recovered recently 
\cite{petrov03} as a particular limit for different
masses of the two fermion species.) 
One may then wonder whether atom-dimer scattering will also show resonance
enhancement,  whether three-body physics remains universal (i.e., determined
by two-body quantities only), and whether a
three-body bound state (`trimer') is
possible. We find that in the fermionic case (with equal masses),
the low-energy physics is universal like in the 3D case \cite{efimov}, and there
is no trimer state. However, in the bosonic case, important
differences to these answers for fermions can and do arise.

At this point, we pause in order to briefly discuss the 
bosonic three-body problem in 3D \cite{hammer}.
It was shown by Thomas in 1935 \cite{thomas} 
that there is no lower bound on the
energy of a system of three bosons interacting via zero-range forces
(`Thomas collapse').
The corresponding integral equation for the scattering amplitude was derived
by STM \cite{skorniakov}, but in Ref.~\cite{danilov} it was shown 
that their equation is not well-defined,
since it allows for an infinite number of solutions.
A scheme of
choosing the correct solution based on the orthogonality of wavefunctions
with different energies was suggested. 
A related equation for the bound states was investigated in Ref.~\cite{minlos}, where an infinite number of 
three-body bound states with arbitrarily low energies was found. Furthermore,
in the unitary limit $a\to\infty$, bound states condense at the
continuum threshold.
The condition of Ref.~\cite{danilov} was identified as a self-adjoint extension
of the involved operator.
Finally, Efimov solved the problem \cite{efimov} by introducing a
three-body short-distance parameter.  Away from the unitary limit,  
there is at most a finite number of bound states (`Efimov states'),
but in the unitary limit a peculiar hierarchy of infinitely many
bound states emerges \cite{efimov,hammer}. This regularization also
affects the scattering solution. To summarize, the 3D integral 
equations for bosons generally require both a long-distance cutoff 
(the scattering length $a$) and a short-distance cutoff. 
Efimov's real-space implementation of the short-distance cutoff is not
very convenient for the confined problem. Fortunately,
Petrov \cite{petrovreg} recently suggested a different and in the
present context more useful regularization based on the
energy dependence of the scattering length,
\begin{equation}\label{Rstar}
a^{-1}\to a^{-1}+R^* m_0 E_c/\hbar^2,
\end{equation}
evaluated at the collision energy $E_c$ of the atom-dimer complex,
i.e., the total energy minus the (kinetic and confinement) energy of the
relative motion.
Here $R^*$ is a parameter related to the sharpness of the Feshbach
resonance, appearing also in the
effective range expansion for the 3D scattering amplitude \cite{landau},
\begin{equation}\label{effrange}
f_{3D}(k) = - \frac{1}{a^{-1}+i k + R^* k^2}. 
\end{equation}
We shall use the regularization (\ref{Rstar}) in our discussion
of the bosonic three-body problem.
Although in strictly 1D  systems, there is no Efimov physics \cite{hammer},
in the quasi-1D case of a confined gas under study here, Efimov states
can and will be relevant in certain limits.

The tunability of the scattering length $a$ in cold atom experiments
is obtained by using Feshbach resonances.  
 For distances on the order of
the interatomic potential, the two-body problem is coupled 
resonantly to a bound molecular state
involving different spin states (closed channel). For larger distances, 
there is nevertheless only a small admixture of this closed channel
and the two-body scattering is essentially reproduced by a one-channel
zero-range potential with the scattering amplitude \eqref{effrange},
 omitting the (small) background scattering length.
We refer to Ref.~\cite{shlyapnikov} for more details.
For fermions, we will assume that $R^*$ does not appear in Eq.~\eqref{effrange}
under the condition that $k R^* \ll 1$. This implies in particular
that $R^* \ll a_B$ and $R^* \ll a_\perp$.

The structure of this paper is as follows. In Sec.~\ref{sec2},
we derive an integral equation that determines the complete
solution of the fermionic three-body problem.  This integral equation
can be solved explictly after projection to the transverse
ground-state of the trap, as is discussed in Sec.~\ref{sectrans}.
The role of the higher transverse channels is then addressed
in Sec.~\ref{secclos}, where also explicit contact to previous
work in the unconfined case \cite{skorniakov,petrov03} is established.
A brief account of some of these results for fermions has been
given in Ref.~\cite{prl}.
The bosonic three-body problem is then studied in detail in Sec.~\ref{secbos}.
In Sec.~\ref{bethe}, we highlight the tight connections to the 
Bethe Ansatz existing in the BCS limit for fermions and bosons, 
and present exact results obtained from this link.
Finally, we conclude in Sec.~\ref{conc}.

\section{Fermionic three-body problem}\label{sec2}

In this section, we consider the fermionic three-body problem
 $(\uparrow\uparrow\downarrow)$ with
two identical fermions, where the `spin' indicates two
selected hyperfine states of the atom.
We denote by ${\bf x_1}$ $({\bf x_{2,3}})$ the 
 position of the $\downarrow$ (the two $\uparrow$) particles, and
perform an orthogonal transformation to variables $({\bf x}, {\bf y}
,{\bf z})$ in order to decouple the center-of-mass coordinate ${\bf z}$,
\begin{equation} \label{orthogonal}
\begin{pmatrix} {\bf x} \\ {\bf y} \\ {\bf z} \end{pmatrix} 
= \begin{pmatrix}
2/\sqrt{3} & -1/\sqrt{3} &  -1/\sqrt{3} \\ 0 & -1 & 1 \\
\sqrt{2/3} &  \sqrt{2/3} &  \sqrt{2/3} \end{pmatrix}
\begin{pmatrix} {\bf x}_1 \\ {\bf x}_2 \\ {\bf x}_3 \end{pmatrix} .
\end{equation}
For the harmonic confinement (\ref{confinment}),
the potential remains diagonal in the positions,
and the Schr\"odinger equation reads
\begin{equation}\label{schro}
 \left( -\frac{\hbar^2}{m_0} \nabla_{\bf X}^2 + U_c({\bf X}) -E \right)
 \Psi({\bf X}) = - \sum_{\pm}  V ( {\bf r}_{\pm} )  \Psi({\bf X}),
\end{equation}
where ${\bf X} = ({\bf x},{\bf y})$ is a six-dimensional vector.
With these definitions, the distances between 
the $\downarrow$ particle and each $\uparrow$ particle 
are
\begin{equation}\label{rpm}
{\bf r}_{\pm} = \sqrt{3}{\bf x}/2 \pm  {\bf y}/2  =  \sin\theta \, {\bf x}\pm
\cos\theta \, {\bf y},
\end{equation}
where we introduce the angle $\theta=\pi/3$ for
notational convenience, see also Ref.~\cite{petrov03}, such that
 $\sin \theta = \sin(2\theta)=\sqrt{3}/2$ and $\cos \theta = 
-\cos(2\theta)=1/2$.
 Equation (\ref{pseudopot})  then
allows to incorporate the interactions $V({\bf r}_\pm)$ via
boundary conditions imposed for vanishing distances between $\uparrow$ and
$\downarrow$ atoms.  For ${\bf r}_{\pm} \to 0$, this implies the
singular behavior
\begin{equation}\label{asymto}
\Psi({\bf X}) \simeq \mp \frac{f({\bf r}_{\perp,\pm})}{4 \pi r_\pm} (1-
r_\pm/a ),
\end{equation}
where the vectors ${\bf r}_{\perp,\pm} = \cos  \theta  \, {\bf x} \mp
\sin \theta \, {\bf y}$ are orthogonal to ${\bf r}_{\pm}$, respectively.
Since the pseudopotential (\ref{pseudopot}) acts on $\Psi$ according to
\[
-\frac{m_0}{\hbar^2}\,
 V( {\bf r}_{\pm} ) \Psi({\bf X}) = \mp f({\bf r}_{\perp,\pm})
\delta( {\bf r}_{\pm} ),
\]
the Schr\"odinger equation (\ref{schro}) becomes
\begin{eqnarray}\nonumber
\left( -\frac{\hbar^2}{m_0} \nabla_{\bf X}^2 + U_c({\bf X}) -E\right)
 \Psi({\bf X}) &=&
\sum_{\pm}  \mp \frac{\hbar^2\,f({\bf r}_{\perp,\pm}) }{m_0}
\delta( {\bf r}_{\pm} ) \\ \label{schro2} &=& S({\bf x},{\bf y}).
\end{eqnarray}
Now we introduce the two-particle Green's function
\begin{equation}\label{greens2}
G_E^{(2)} ( {\bf r}_1,{\bf r}_2; {\bf r}_3, {\bf r}_4) 
= \sum_{\lambda_1,\lambda_2} \frac{\psi_{\lambda_1} (  {\bf r}_1 )
\psi_{\lambda_2} (  {\bf r}_2 ) \psi_{\lambda_1}^* (  {\bf r}_3 )
\psi_{\lambda_2}^* (  {\bf r}_4 )}{E_{\lambda_1} + E_{\lambda_2} -  E},
\end{equation}
where $\psi_\lambda$ denotes the eigenfunctions to the single-particle
problem (for reduced mass $m_0/2$) with eigenenergy $E_\lambda$. The quantum
numbers $\lambda$ include the longitudinal (1D) momentum $k$, the integer
angular momentum $m$,
and the radial quantum number $n=0,1,2\ldots$, whence 
\begin{equation}\label{eigen}
\begin{split}
E_\lambda &= \hbar \omega_\perp(2n+|m|+1) + \hbar^2 k^2/m_0 \\ 
\psi_\lambda& = e^{im\phi}R_{nm}(\rho)e^{ikz} 
\end{split}
\end{equation}
with radial functions [see also Eq.~(\ref{aperp})] 
\begin{eqnarray}\label{radial}
R_{nm}(\rho) &=& \frac{1}{\sqrt{\pi}\ a_\perp} \sqrt{n!/(n+|m|)!}
\\ \nonumber
&\times& (\rho/a_\perp)^{|m|} L_n^{|m|} (\rho^2/a^2_\perp)
e^{-\frac12 (\rho/a_\perp)^2} ,
\end{eqnarray}
where $L_n^m(x)$  denotes standard Laguerre polynomials.
Using Eq.~(\ref{greens2}), the general solution of
Eq.~\eqref{schro2} can be expressed as
\begin{equation}\label{gen2}
\Psi({\bf X}) = \Psi_0({\bf X}) + \int d {\bf x'} d {\bf y'}
G_E^{(2)} ({\bf x},{\bf y};{\bf x}',{\bf y}') S({\bf x}',{\bf y}'),
\end{equation}
where $\Psi_0({\bf X})$, only present for positive energy $E$,
is a homogeneous (free) solution.
In this paper, we restrict our attention to states with $E<0$, 
such that $\Psi_0=0$, and write 
\begin{equation}\label{energy1}
E = -2 \Omega_B \, \hbar \omega_\perp + \hbar^2 \bar{k}^2/m_0,
\end{equation}
where the relative momentum $\bar{k}$ of the atom-dimer complex
is sent to zero later.

Since $S({\bf x},{\bf y})$ in Eq.~(\ref{gen2}) 
involves the variables ${\bf r}_{\perp,\pm}$
and ${\bf r}_{\pm}$, see Eq.~(\ref{schro2}),
it is advantageous to switch to these 
by virtue of the orthogonal transformation
\begin{equation}\label{ortho1}
\begin{pmatrix}
 {\bf r}_{\pm} \\ {\bf r}_{\perp,\pm} \end{pmatrix} = \begin{pmatrix}
\sin \theta & \pm \cos \theta \\ \cos \theta & \mp \sin \theta \end{pmatrix}
\begin{pmatrix}
{\bf x} \\ {\bf y} \end{pmatrix}.
\end{equation}
Similarly, we can switch from $({\bf r}_{\perp,-},{\bf r}_{-})$  to
$({\bf r}_{\perp,+},{\bf r}_{+})$ using
\begin{equation}\label{changebasis}
\begin{pmatrix} {\bf r}_{+} \\ {\bf r}_{\perp,+} \end{pmatrix} = 
\begin{pmatrix} - \cos 2  \theta & \sin 2 \theta \\ 
\sin 2 \theta & \cos 2 \theta \end{pmatrix} \begin{pmatrix}
 {\bf r}_{-} \\ {\bf r}_{\perp,-} \end{pmatrix}.
\end{equation}
Since these are orthogonal transformations, $G_E^{(2)}$ stays invariant,
\[
G_E^{(2)} ({\bf x},{\bf y};{\bf x}',{\bf y}') = G_E^{(2)} (
{\bf r}_{\pm}, {\bf r}_{\perp,\pm} ;  {\bf r}_{\pm}', {\bf r}_{\perp,\pm}').
\]
Furthermore, the integration measure in Eq.~(\ref{gen2}) is
also invariant,
$d  {\bf x'} d {\bf y'} = d {\bf r}_{\perp,\pm}' d {\bf r}_{\pm}'.$
We thus find from Eq.~(\ref{gen2})
\begin{eqnarray}\label{final}
&& \frac{m_0}{\hbar^2}\,
 \Psi ({\bf r},{\bf r}_\perp) =  \int d {\bf r}^\prime_\perp  \,
f({\bf r}_\perp^\prime) \Bigl [
G_E^{(2)}( {\bf r}  , {\bf r}_\perp; 0,{\bf r}_\perp^\prime)  \\ \nonumber
&-& G_E^{(2)} \Bigl (-\cos(2\theta){\bf r} + \sin(2\theta) {\bf r}_\perp  , 
\sin(2\theta) {\bf r}
+\cos(2\theta) {\bf r}_\perp; 0 ,{\bf r}^\prime_\perp \Bigr) \Bigr ],
\end{eqnarray}
where ${\bf r} \equiv  {\bf r}_{-}$ and ${\bf r}_\perp
\equiv {\bf r}_{\perp,-}$.
Next we implement the ${\bf r} \to 0$ limit according to Eq.~(\ref{asymto})
to obtain a closed equation for $f({\bf r}_{\perp})$.
This limit can be directly taken for the non-singular second term 
in Eq.~(\ref{final}),
while the first term contains the  singular behavior necessary from
 Eq.~(\ref{asymto}).
Once this singular behavior is removed, one obtains a regular
integral equation for $f({\bf r}_{\perp})$ 
\cite{petrov03,skorniakov}.

It is then convenient to
transform into the complete basis $\{ \psi_\lambda \}$ specified above,
$f({\bf r}_\perp)=\sum_\lambda f_\lambda \psi_\lambda({\bf r}_\perp)$.
Notably, the last term in Eq.~(\ref{final}) can 
be expressed in terms of the single-particle Green's function 
\begin{equation}\label{green1}
G_E({\bf r},{\bf r}') = \sum_\lambda \frac{\psi_\lambda({\bf r})
\psi_\lambda^*({\bf r}')}{E_\lambda-E} ,
\end{equation}
since 
\begin{eqnarray*}
 && \int d {\bf r}'_\perp
\ G_E^{(2)}( \sin (2 \theta) {\bf r}_\perp,   \cos (2 \theta) \,
{\bf r}_\perp ; 0,{\bf r}'_\perp) f({\bf r}'_\perp) = \\ &&
\sum_\lambda G_{E-E_\lambda} (  \sin (2 \theta) \, {\bf r}_\perp,0 )
\ \psi_{\lambda} (  \cos (2 \theta) \,{\bf r}_\perp ) \ f_\lambda.
\end{eqnarray*}
The Green's function (\ref{green1}) for ${\bf r}'=0$
has the integral representation
\begin{eqnarray}\label{tokat}
G_E({\bf r},0) &=& 
\frac{m_0}{4\pi \hbar^2 a_\perp}\int_0^\infty\frac{ dt}{\sqrt{\pi t}}
\frac{e^{-\Omega t}}{1-e^{-t}} \\
\nonumber &\times& \exp\left(-\frac{z^2}{a_\perp^2 t} - 
\frac{\rho^2}{2a_\perp^2} \coth(t/2)\right),
\end{eqnarray}
where $\Omega= (\hbar \omega_\perp-E)/2 \hbar \omega_\perp$.
This can be obtained from the 
Feynman (imaginary-time) representation of 
the Green's function (\ref{green1}), 
\[
G_E({\bf r},{\bf r'}) 
= \int_0^{\infty} d t \sum_\lambda \psi_\lambda({\bf r})
\psi_\lambda^*({\bf r}') e^{-t (E_\lambda-E)},
\]
and the expressions for the 
eigenfunctions $\{ \psi_\lambda \}$ and eigenenergies $E_\lambda$,
see Eqs.~\eqref{eigen} and \eqref{radial}, where only states with
$m=0$ contribute. One has to perform a straightforward
gaussian integral over momenta $k$ and the sum over $n$ follows from 
the remarkable identity
\[
\sum_{n=0}^{\infty} L_n^0 \left( \frac{\rho^2}{a_\perp^2} \right)
e^{-2 n \hbar \omega_\perp t} = \frac{1}{1-e^{-2 \hbar \omega_\perp t}}
\exp \left( \frac{\rho^2}{a_\perp^2} \frac{e^{-2 \hbar \omega_\perp t}}
{e^{-2 \hbar \omega_\perp t}-1} \right),
\]
leading finally to Eq.~\eqref{tokat}.
Using the orthonormality of the $\{ \psi_\lambda \}$,
the first term in the integral appearing in Eq.~\eqref{final} can be written
as
\begin{eqnarray}\label{decomp}
&& \int d {\bf r}'_\perp G_E^{(2)}( {\bf r}  , {\bf r}_\perp; 0,{\bf r}'_\perp) 
f({\bf r}'_\perp) \\ \nonumber
&& = \sum_{\lambda,\lambda_1} \frac{\psi_{\lambda_1} ({\bf r})
\psi_{\lambda_1}^*
(0) }{E_{\lambda_1} + E_\lambda - E} \psi_{\lambda} ( {\bf r}_\perp )
f_\lambda \\ \nonumber
&&= \sum_{\lambda} G_{E-E_\lambda} ({\bf r},0)
\psi_{\lambda} ( {\bf r}_\perp ) f_\lambda.
\end{eqnarray}
For ${\bf r}\to 0$, the integral representation in Eq.~(\ref{tokat})
is dominated by small values of $t$. Its asymptotic behaviour is then
obtained by substracting and adding the leading term, such that
\[
\begin{split}
G_E({\bf r},0) = \frac{m_0}{4\pi \hbar^2 a_\perp}
\left( \int_0^\infty\frac{ dt}{\sqrt{\pi} t^{3/2}}
e^{-\Omega t - r^2/ a_\perp^2 t} \right.\\[1mm]
 + \zeta(1/2,\Omega) + o(1) \bigg),
\end{split}
\]
where the integral representation of 
the Hurvitz zeta function \eqref{zeta} has been used.
Finally, performing the $t$ integral, we find that for ${\bf r}\to 0$,
Eq.~(\ref{tokat}) has the asymptotic
behaviour  $G_E ({\bf r},0) \simeq (m_0/4\pi \hbar^2 a_\perp) [a_\perp/r
+ \zeta(1/2,\Omega)]$.
This implies that the leading term in Eq.~\eqref{decomp} gives
$m_0 f( {\bf r}_\perp ) / (4 \pi \hbar^2 r)$, which coincides with
the singular part of $m_0  \Psi ( {\bf r} ,  {\bf r}_\perp)/\hbar^2$
in the  $r \to0$ limit.

Using Eq.~(\ref{asymto}), and
cancelling the $r\to 0$ singular terms in Eq.~\eqref{final}, 
straightforward algebra leads to an {\sl integral equation} for $f({\bf r})$.
In the $\{ \psi_\lambda \}$ representation, it reads
\begin{equation}\label{integral-equa}
{\cal L}(\Omega_\lambda) f_\lambda = 
\sum_{\lambda'} A_{\lambda,\lambda'} \, f_{\lambda'},
\end{equation}
where we use the function, see also Eqs.~\eqref{zeta} and
\eqref{bs},
\begin{equation} \label{lome}
\mathcal{L}(\Omega)  =  \zeta(1/2, \Omega) - \zeta(1/2,\Omega_B) ,
\end{equation}
and the frequencies
\begin{equation}\label{omegalam} 
 \Omega_\lambda  = \Omega_B - (a_\perp \bar k/2)^2
+ E_\lambda/2 \hbar \omega_\perp.
\end{equation}
We find the kernel in the form
\begin{eqnarray} \nonumber
A_{\lambda\lambda'}&=&\frac{4\pi \hbar^2 a_\perp}{m_0}
\int d {\bf r}_\perp \, \psi_{\lambda}^* ( {\bf r}_\perp)\,
\psi_{\lambda'} ( \cos(2\theta) {\bf r}_\perp ) \\  &\times&
\label{lomega}
G_{E-E_{\lambda^\prime}} (  \sin(2\theta) {\bf r}_\perp,0 ).
\end{eqnarray}
 In the following, it will be more convenient to use $\Omega_B$ instead of 
$a/a_\perp$ to parametrize the interaction strength, see Eq.~(\ref{bs}).

Using Eq.~(\ref{tokat}), $A_{\lambda,\lambda'}$ can be evaluated
explicitly, but before proceeding further, 
we shall perform a rescaling.  
Until now,  $f$ has only been considered as a function of
the variable ${\bf r}_\perp = (\rho,z)$.
However, in the asymptotic three-body scattering solution consisting
of a dimer and one unbound atom, the atom-dimer distance ${\bf d}$
(which is then much bigger than the dimer size $a_B$)
coincides with ${\bf r}_\perp$ only after a proper rescaling. 
The asymptotic solution is expected to be of the form
\begin{equation}\label{asymto1}
\Psi ( {\bf r} ,  {\bf r}_\perp) = \Phi_0 ( {\bf r} ) \chi ({\bf d}),
\end{equation}
where ${\bf r}\equiv {\bf r}_{-}$ and the atom-dimer distance is
\[
{\bf d} \equiv ({\bf x_1} + {\bf x_3})/2 -
{\bf x_2} = {\bf r}_+ - {\bf r}_- /2.
\]
Here $\Phi_0 ( {\bf r} )$ is the wavefunction of the 
confinement-induced two-body bound state \cite{olshanii98},
and $\chi ({\bf d})$ gives the asymptotic solution for the scattering of
 the free particle by the dimer.
The connection with $f$ is made by looking at the ${\bf r} \to 0$
limit of Eq.~\eqref{asymto1}. The leading term is
\[
\Psi ( {\bf r} ,  {\bf r}_\perp) \simeq \frac{1}{4 \pi r} \,
\chi ( \sin (2 \theta)   {\bf r}_\perp ),
\]
where we have used Eq.~\eqref{changebasis} to express ${\bf r}_+$
as a function of ${\bf r}$ and ${\bf r}_\perp$.
In the asymptotic limit, $\sin (2 \theta)  {\bf r}_\perp$
is thus the atom-dimer distance.
Therefore, after the rescaling ${\bf r}_\perp \to \sin (2 \theta) 
{\bf r}_\perp$, the function $f$ matches
the asymptotic scattering solution $\chi$.
This length rescaling also implies 
wavevector rescaling,
$k  \to k / \sin (2 \theta)$, as well as an extra factor $\sin(2\theta)$ in
$A_{\lambda, \lambda'}$.
In addition, from now on, 
we switch to dimensionless lengths and wavevectors by
measuring them in units of $a_\perp$ and $1/a_\perp$, respectively.

\section{Lowest transverse channel}
\label{sectrans}

Let us then proceed by projecting the integral equation
 (\ref{integral-equa}) to the lowest transverse state ($n=m=0$).
The role of the higher transverse channels will be discussed
in Sec.~\ref{secclos}.
Taking into account the above rescaling, and noting that only $m=0$
modes have nonzero overlap with the lowest state,
\begin{equation}\label{equa}
{\cal L}(\Omega_k) f_k 
= \int_{-\infty}^{\infty} \frac{d k'}{2 \pi}  A_{k,k'}  f_{k'},
\end{equation}
where 
\[
\Omega_k =\Omega_B +\sin^2(2\theta) (k^2-\bar{k}^2)/4.
\]
Straightforward algebra gives
\begin{eqnarray}\label{deltakk}
&& A_{k,k'}  =  \sum_{p=0}^{\infty}  \left(\frac{1+\cos(4\theta)}{2}\right)^p
 \\ \nonumber && \times \frac{1}{p+ \Omega_B 
+ [k^2 + k'^2 + k k']/4 -3 \bar{k}^2/16} .
\end{eqnarray}
Note that the energy reads after the rescaling
\begin{equation}\label{energy}
E= -2 \hbar \omega_\perp \Omega_B + \frac{3}{4} \frac{\hbar^2}{m_0} \,
\left(\frac{\bar{k}}{a_\perp}\right)^2,
\end{equation}
where $\bar{k}$ is interpreted as the relative momentum
of a free particle with reduced mass $2m_0/3$.

\subsection{Atom-dimer scattering solution}\label{subsecA}

Following STM
\cite{skorniakov}, we now make an {\sl Ansatz} for the
solution of this integral equation, 
\begin{equation}\label{skor}
f (k) = 2 \pi \delta( k-\bar{k} ) + i \tilde{f} (k,\bar{k} ) \sum_\pm
\frac{1}{\bar{k} \pm k + i 0^+}, 
\end{equation}
with a {\sl regular} function (scattering amplitude) 
$\tilde{f}(k,\bar{k})$. This
Ansatz gives the expected asymptotic scattering state after Fourier 
transforming to real space, 
\[
f(z) = e^{i\bar{k} z}  +  \tilde{ f}
\left ( {\rm sgn}(z) \bar{k}, \bar{k}\right ) e^{i\bar{k} |z|}, \qquad
|z| \to +\infty,
\]
such that standard transmission and  reflection amplitudes \cite{landau}
can be inferred,
\begin{equation}\label{refl}
t(\bar{k}) = 1+ \tilde{f}(\bar{k},\bar{k}) ,\quad 
r(\bar{k})= \tilde{f}(-\bar{k},\bar{k}).
\end{equation}
In the low-energy limit $k,\bar{k}\to 0$, the general expansion of the
scattering amplitude applies,
\begin{equation} \label{aaddef}
\tilde{f}(k,\bar{k}) =
 -1 + i k b_{ad} + i\bar{k} a_{ad} + {\cal O}(k^2, \bar{k}^2,
k \bar k).
\end{equation}
Here $a_{ad}$ and $b_{ad}$ are the two 1D scattering lengths
for the {\sl atom-dimer scattering} process.  In general,
$a_{ad}$ and $b_{ad}$ are related to the even and odd
partial scattered waves, respectively, as will be discussed
in more detail in Sec.~\ref{bosonic}. 
Let us just note at this point that for a sufficiently short-ranged
potential, namely with support $a_s$ smaller than the typical
lengthscale for the wavefunction variations, the potential can
be effectively described by a contact $\delta$-interaction.
This requires $a_s \ll |a_{ad}|$ and $\bar{k} \ll 1/a_s$.
In that case, odd waves are not scattered by the potential 
and hence $b_{ad}=0$ from Eq.~\eqref{aaddef2}. This is the case
in particular for the two-body problem where only one scattering
length, $a_{aa}$ in Eq.~\eqref{a1d}, is usually given.

Inserting the Ansatz (\ref{skor}) into Eq.~(\ref{equa}), we obtain
\begin{eqnarray}
\nonumber
&& \frac{\mathcal{L} ( \Omega_{k} )}{\bar{k}^2 - k^2} 2 i \bar{k}
\tilde{f}(k,\bar{k}) - i \mathcal{P} \sum_\pm
\int_{-\infty}^{\infty} \frac{d k'}{2 \pi}
 \frac{A_{k,k'}}{\bar{k} \pm k'}  
 \tilde{f} (k',\bar{k}) \\  
\label{eqint2}
&& - \frac12 \left[ \tilde{f} (\bar{k},\bar{k})   A_{k,\bar{k}}+ 
\tilde{f} (-\bar{k},\bar{k})   A_{k,-\bar{k}} \right] = A_{k,\bar{k}},
\end{eqnarray}
where ${\cal P}$ denotes a principal value integration.
One can check that
the function $\mathcal{L}(\Omega_k) / ( \bar{k}^2 - k^2)$ is regular
when $\bar{k} \to k$, and we used above that $\mathcal{L}(
\Omega_{\bar{k}})=0$.
Equation (\ref{eqint2}) is an inhomogeneous integral equation of
the second kind for a given value of $\bar{k}$, which has
 a unique solution if the corresponding kernel is invertible.
In principle, it may be solved numerically for any value of $\bar{k}$
to extract the value of $a_{ad}$.  

However, the subsequent analysis is simplified considerably
by letting $\bar{k} \to 0$.  Formally, we expand Eq.~(\ref{eqint2})
in $\bar{k}$ and keep only the lowest order. 
To that purpose, we first rewrite Eq.~\eqref{eqint2} in the form
\begin{eqnarray*}
&&
\frac{\mathcal{L} ( \Omega_{k} )}{\bar{k}^2 - k^2} 2 i \bar{k}
\tilde{f}(k,\bar{k}) + i \mathcal{P} 
\int_{-\infty}^{\infty} \frac{d k'}{2 \pi k'}
\Bigl[  \tilde{f} (k'+\bar{k} ,\bar{k}) \, A_{k,k'+\bar{k}} \\
&& -
  \tilde{ f} (k'-\bar{k} ,\bar{k})  \, A_{k,k'-\bar{k}} \Bigr] \\
&& - \frac 1 2 \left[ \tilde{f} (\bar{k},\bar{k})   A_{k,\bar{k}}+
\tilde{f} (-\bar{k},\bar{k})   A_{k,-\bar{k}} \right] = A_{k,\bar{k}}.
\end{eqnarray*}
Expanding in $\bar{k}$ and dividing by $2 \bar{k}$,
we then get to lowest order:
\begin{equation}\label{eqint3}
\begin{split}
- i \frac{\mathcal{L} ( \Omega_{k} )}{k^2} &
\tilde{f}(k,0)  + i \mathcal{P} \int_{-\infty}^{\infty} \frac{d k'}{2 \pi k'}
\partial_{k'} \left[ A_{k,k'} \tilde{f}(k',0) \right] \\[3mm]
& = A_{k,0} \, \left( \frac{\tilde{f}(0,\bar{k})+1}{2 \bar{k}}
\right)_{\bar{k} \to 0} + \frac 1 2 \, \partial_{k'} A_{k,k'=0}.
\end{split}
\end{equation}
Finally, we integrate by parts,
use Eqs.~(\ref{deltakk}) and (\ref{aaddef}),  
and switch to dimensionless momenta  by writing
$k=2\sqrt{\Omega_B} u$.  Collecting terms, we  then arrive at a 
tractable integral equation for  $h(u)\equiv \tilde{f}(u,0)$.
With the weakly $\Omega_B$-dependent functions
\begin{eqnarray}\label{Guu}
G(u,u')&=& \sum_{p=0}^\infty \frac{4^{-p}}{1+u^2+u^{\prime 2}+ 
u u^\prime + p/\Omega_B},\\ \label{hu}
H(u) &=& \sum_{p=0}^\infty \frac{4^{-p} u}{2(1+u^2+p/\Omega_B)^{-2} },
\end{eqnarray}
this integral equation  reads
\begin{eqnarray} \nonumber
&& \int_{-\infty}^{\infty} \frac{d u'}{ 2\pi u'^2} 
[ G(u,u') h(u') -  G(u,0) h(0)]  
\\ \label{inteqfin}
&& - \frac{ \sqrt{\Omega_B}}{2u^2} 
{\cal L}\left(\Omega_B\left[1+\frac{3u^2}{4}\right]\right) h(u)
\\ \nonumber &&= \frac{a_{ad}\sqrt{\Omega_B}}{a_\perp} G(u,0) +iH(u). 
\end{eqnarray}
Note that the real (imaginary) part of $h(u)$ is even (odd) in $u$.
The scattering length $a_{ad}$ finally follows from 
the real part of Eq.~(\ref{inteqfin}) and the condition $h(0)=-1$,
see Eq.~(\ref{aaddef}), while $b_{ad}$ can be extracted from the
imaginary part of Eq.~(\ref{inteqfin}).
The integral equation (\ref{inteqfin})
 shows in particular that $a_{ad}/a_\perp$ and $b_{ad}/a_\perp$ depend
 only on $\Omega_B$, and hence only on the binding energy of the dimer.
This already suggests universality of the three-fermion problem.

\subsection{Dimer limit}

The solution of this integral equation is discussed first for  
 $a_\perp/a\gg 1$, where
 tightly bound dimers of size $a_B\approx a$  and large binding
energy, $\Omega_B=(a_\perp/2a)^2\gg 1$, are realized.
Expanding the real part of Eq.~(\ref{inteqfin}) in $1/\Omega_B$,
carefully including the $\Omega_B$-dependence of $G(u,0)$ and ${\cal L}$,
and using $\zeta(1/2,\Omega\gg 1)\approx -2\sqrt{\Omega}$,
we obtain to first order 
\begin{eqnarray*}
&& \frac 3 4  \frac{h(u)}{1+\sqrt{1+3u^2/4}} 
+ \frac{3}{16} \frac{1}{\Omega_B}
\frac{h_0(u)}{1+ 3 u^2 /4 + \sqrt{1+3u^2/4}}\\
&& + \frac 4 3 \frac{1}{\Omega_B}
\int_{-\infty}^{\infty} \frac{d u'}{ 2\pi u'^2} \left(
\frac{h_0(u')}{1+u^2+u'^2+u u'} + \frac{1}{1+u^2} \right) \\
&& =  \left( \frac{a_{ad}}{a_\perp \sqrt{\Omega_B}} \right) \left(
\frac 4 3 \frac{1}{1+u^2} - \frac{1}{\Omega_B} \frac{W_0}{(1+u^2)^2} \right)
\end{eqnarray*}
with $G(0,0)=4/3$ and $W_0 = \sum_p p (1/4)^p = 4/9$. Here, we
use the zeroth-order
approximation $h_0(u)$ to the full solution $h(u)$. 
The lowest order gives 
\begin{equation}\label{h0}
h_0(u) = -\frac12 \frac{1+\sqrt{1+3u^2/4}}{1+u^2},
\end{equation}
where $a_{ad}=-(9/32)\sqrt{\Omega_B}$ fixes $h_0(0)=-1$. 
The next order gives from $h(0)=-1$ a correction to the atom-dimer
scattering length, such that
\begin{eqnarray} \label{dimer1}
a_{ad} &=& - \kappa_\infty a_\perp \sqrt{\Omega_B} + \beta a_\perp
 /\sqrt{\Omega_B} \\
\nonumber &=&  - \kappa_\infty a_\perp^2/2a  + 2 \beta a,
\end{eqnarray}
where $\kappa_\infty=9/32=0.28125$ and 
\begin{equation}\label{beta}
\beta= -\frac{9}{128}+\frac{3\sqrt{3}+4\pi}{8\pi}-\frac{3}{32} \simeq 0.5426.
\end{equation}
Similarly, we can then compute $h(u)$ to first order in $1/\Omega_B$,
and thereby obtain $b_{ad}$ from the imaginary
part of Eq.~(\ref{inteqfin}).  
Straightforward algebra gives 
\begin{equation}\label{bad}
b_{ad}/a_\perp = (8/9)\Omega^{-3/2}_B
\end{equation}
for $\Omega_B\gg 1$. The vanishing value of $b_{ad}$ is a consequence
of the short-rangedness of the effective atom-dimer potential \cite{prl}.
The support of this potential is the dimer size $a_B\simeq a$ and goes to 
zero in the dimer limit. This validates a repulsive zero-range 1D
atom-dimer potential in the low-energy limit,
$V_{ad}(z)=g_{ad} \delta(z)$ with $g_{ad}\propto (-1/a_{ad})$,
very similar to the 1D atom-atom scattering potential (\ref{vaa}).
In cold atom systems, the validity of our treatment in the dimer
limit is always limited by the constraint that the 3D scattering 
length $a$ is larger than the typical size of the actual atom-atom
potential.

\subsection{Numerical solution}\label{numerical}

Outside the dimer limit, in general a numerical solution
of Eq.~(\ref{inteqfin}) is necessary. 
We describe next how an accurate numerical solution to 
Eq.~(\ref{inteqfin}) can be obtained in practice.  One has to be quite 
careful to ensure regularity of $h(u)$, for which we found it
beneficial to Fourier transform to real space,
where the Fourier transformed $h$ is well-behaved and allows for 
a quickly converging solution of the integral equation. 
In order to implement an efficient, fast and reliable numerical solution,
it is mandatory to find a convenient representation of the 
integral kernel. As this is a nontrivial problem,
 we outline its solution in some detail here.

Let us first give some auxiliary relations that will be useful below.
The function ${\cal L}$ appearing in Eq.~\eqref{inteqfin} can be written as 
\begin{eqnarray*}
{\cal L}(u)&=& \zeta(1/2,\Omega_B(1+3u^2/4))-\zeta(1/2,\Omega_B)\\
&=& \frac{1}{\sqrt{\Omega_B}}
\int_0^\infty \frac{dt}{\sqrt{\pi t}} \frac{e^{-t}}{1-e^{-t/\Omega_B}}
\left(e^{-\frac{3}{4}u^2 t}-1\right).
\end{eqnarray*}
In addition, $G(u,u')$ in Eq.~(\ref{Guu}) can alternatively be expressed in
the form
\[
G_{\Omega_B}(u,u')= \int_0^\infty \frac{dt}{1-\frac{1}{4}
e^{-t/\Omega_B}}e^{-(1+u^2
+uu' + u^{\prime 2})t}.
\]
We then switch back to real space by writing
\begin{equation}
h(u) = -\int_{-\infty}^\infty dz e^{izu} g(z),
\end{equation}
which leads to the Fourier transform of Eq.~(\ref{inteqfin}).
The real part of this equation is
\begin{equation}\label{realspace}
\int_{-\infty}^\infty
 dz' K(z,z') g(z') = -\frac{ a_{ad}\sqrt{\Omega_B}}{a_\perp} B(z)
\end{equation}
with
\begin{eqnarray} \nonumber
K(z,z') &=&
\int \frac{du}{2\pi} \int \frac{du'}{2\pi u^{\prime 2}}
e^{-izu} ( G(u,u')e^{iz'u'} - G(u,0)) \\ \label{kdef0}
&-& \frac{\sqrt{\Omega_B}}{2} \int \frac{du}{2\pi u^2}
 {\cal L}(u) \ e^{-i(z-z')u} \\ \nonumber
& =& K_1(z,z')+ K_2(z-z') \\ \nonumber
B(z) &=&
\int\frac{du}{2\pi}e^{-izu}
G(u,0) = \sum_{p=0}^\infty 4^{-p} \frac{e^{-\sqrt{1+p/\Omega_B}|z|}}{
2\sqrt{1+p/\Omega_B}}.
\end{eqnarray}
Using the above auxiliary relations,
some algebra gives with $X(t)\equiv |z'+z/2|/\sqrt{3t}$ and
the probability function $\Phi(X)$ a convenient representation for
the function $K_1$,
\begin{eqnarray}\label{kdef1}
K_1(z,z') &=& - \frac{\sqrt{3}}{4\sqrt{\pi}} \int_0^\infty dt
 \frac{e^{-t} e^{-z^2/4t}}{1-\frac{1}{4}e^{-t/\Omega_B}} \\
&\times& \nonumber
\left[ X(t) \Phi(X(t)) + \frac{1}{\sqrt{\pi}} e^{-X^2(t)} \right].
\end{eqnarray}
In a similar fashion, with $Z(t) = |z-z'|/\sqrt{3 t}$, we find 
\begin{equation}\label{kdef2}
\begin{split}
K_2 (z-z') & = \frac 1 4 \sqrt{\frac{3}{\pi}}
\int_0^{\infty} \frac{d t \, e^{-t}}{1-e^{-t/\Omega_B}} \\[2mm]
& \times
\left( \frac{e^{-Z(t)^2}}{\sqrt{\pi}} + Z(t)  [ \Phi (Z(t)) -1 ] \right).
\end{split}
\end{equation}
With $K(z,z^\prime)=K_1+K_2$ and $B(z)$ given, one numerically computes $g(z)$ 
from Eq.~(\ref{realspace}) and then
fixes $a_{ad}$  from the normalization 
condition $\int dz g(z)=1$ corresponding
to $h(0)=-1$.
 
Let us first discuss this program in the BCS limit, $\Omega_B\ll 1$.
Importantly, the kernel $K(z,z')$
is {\sl not invertible} in this limit, since there is a zero mode 
solution $g_0(z)$. This function can be found analytically 
in momentum space using the Bethe Ansatz as described in Sec.~\ref{fermionic},
with the result $h_0(u) = u/(1+u^2)$, leading to the real-space form
$g_0(z) = {\rm sgn} (z) e^{-|z|}$.
Fortunately, the zero mode does not affect
the determination of the scattering length $a_{ad}$.
To see this,  note that $g_0 (z)$ has odd parity and hence does
not contribute to the normalization
condition $\int dz \, g(z)=1$.
Since also the function $B(z)$
in Eq.~\eqref{realspace} is even,
we can restrict the
numerical solution to even functions $g(z)$.  On this 
space, $K(z,z')$ is invertible even in the BCS limit,
and hence the numerical procedure is stable and reliable.

\begin{figure}[t!]
\begin{center}
\includegraphics[scale=0.3]{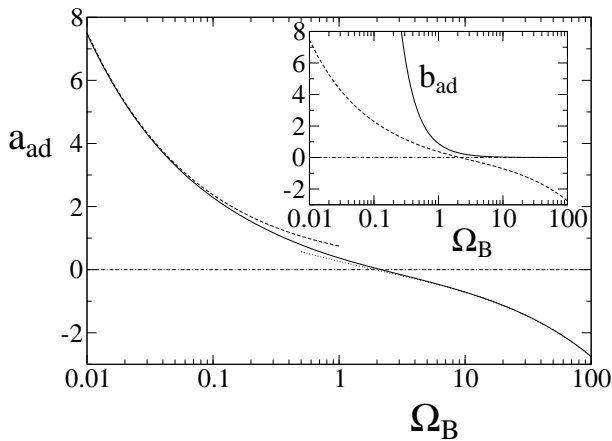} 
\caption{\label{fig2} Scattering length $a_{ad}/a_\perp$ versus dimensionless
binding energy $\Omega_B$. The solid curve is the numerical solution to
Eq.~(\ref{inteqfin}), and the dotted (dashed) curves represent the 
analytical results in the dimer (BCS) limit, respectively. 
In the inset, $b_{ad}/a_\perp$ is plotted
versus $\Omega_B$. The solid curve gives $b_{ad}$ from the numerical 
solution of Eq.~(\ref{inteqfin}), and the dashed line represents $a_{ad}$
for comparison.
}
\end{center}
\end{figure}

The numerical result for $a_{ad}/a_\perp$ as function
of $\Omega_B$, covering the full crossover, is shown in Figure \ref{fig2}.
In the BCS limit, where $a_\perp/a\ll -1$ and $\Omega_B\ll 1$, 
the real part of Eq.~\eqref{inteqfin} is solved by $h(u) = -1/(1+u^2)^2$,
see Eq.~\eqref{resh} below, and hence we find the exact result 
\begin{equation}\label{bcs1}
a_{ad} = \frac{3}{4} \, a_\perp /\sqrt{\Omega_B} = 0.75 \,a_\perp^2/|a|.
\end{equation}
The numerical solution shown in Fig.~\ref{fig2} for arbitrary $\Omega_B$
nicely matches onto the analytically available
limits (\ref{dimer1}) and (\ref{bcs1}).
Remarkably, around $\Omega_B\approx 2.2$, there is a zero of
the atom-dimer scattering length $a_{ad}$.
One may suspect that this behavior is the analogue of the 
two-body CIR in Eq.~(\ref{a1d}). However, first one may 
notice that the atom-dimer `resonance'
occurs at a different $\Omega_B$ than the CIR. 
In addition, it is important to check  whether one
still has a simple contact $\delta$-interaction potential,
since otherwise the simple relationship $g_{ad}\propto -1/a_{ad}$ 
breaks down. 

Let us then address the numerical evaluation of $b_{ad}$ in
Eq.~(\ref{aaddef}), which is performed by looking
at the imaginary part corresponding to
Eq.~\eqref{realspace},
\begin{equation}\label{kerbad}
\int_{-\infty}^\infty dz' K(z,z') g(z') = -B_2 (z),
\end{equation}
where
\begin{eqnarray*}
B_2 (z) & = & \int \frac{d u }{2\pi} e^{-i z u} \, u \, H(u)\\
&=& -i \, \frac{z}{8} \, \sum_{p=0}^{+\infty} \left( \frac{1}{4} \right)^p
\frac{e^{-\sqrt{1+p/\Omega_B}|z|}}{\sqrt{1+p/\Omega_B}}.
\end{eqnarray*}
Once this integral equation is solved, $b_{ad}$ follows as
\begin{equation}
\frac{b_{ad}}{a_\perp} = \frac{1}{2 \sqrt{\Omega_B}}
\int_{-\infty}^{\infty} dz \, z \left[ -i g(z) \right].
\end{equation}
Even though the kernel in Eq.~\eqref{kerbad}
is the same as for the computation of $a_{ad}$, the situation
is quite different in the BCS limit.
The  function $B_2(z)$ is an odd function of $z$
and therefore not orthogonal to $g_0(z)$.
This implies that the integral equation
\eqref{kerbad} has no solution 
when $\Omega_B$ is taken directly to zero.
However, for any finite $\Omega_B$, the
kernel $K(z,z')$ becomes invertible, but the corresponding
solution of \eqref{kerbad} diverges for $\Omega_B \to 0$.
The numerical result for $b_{ad}$ is shown in the inset 
of Fig.~\ref{fig2}. While in the dimer limit, $b_{ad}$ stays small,
in accordance with our analytical result (\ref{bad}), in the 
BCS limit, it is found to diverge as $b_{ad}\propto \Omega_B^{-3/2}$. 
We will see in Sec.~\ref{bethe} that this divergence
originates from the reflectionless scattering
property encountered in the BCS limit.
Moreover, these results for $b_{ad}$ imply that one cannot
use an effective $\delta$-potential for the atom-dimer
scattering outside the dimer limit.
Notably, the vanishing of $a_{ad}$ for $\Omega_B \simeq 2.2$ does
not correspond to an atom-dimer resonance since 
$g_{ad}\propto (-1/a_{ad})$ breaks down away from the dimer limit.

\section{Role of higher transverse channels}\label{secclos}

So far, the role of transverse excited states
($n\ne 0$) has been taken into account only through the exact 
calculation of $A_{k,k'}$ for $n=m=0$. 
We still need to investigate the full Eq.~\eqref{integral-equa}
including higher transverse channels.
In this section, we
address their effect in detail.  For the lowest channel $n=0$,
we keep using the Ansatz (\ref{skor}) for $f(k)$. 
For the higher channels, we have functions $f_n(k)$ instead.
The full integral equation (\ref{integral-equa}) [taking into account the 
rescaling discussed above] leads to a system of coupled integral
equations for $f(k)$ and the $f_{n}(k)$.  For the 
limit $\bar{k}\to 0$, following the line of reasoning in the
last section, we arrive at the previous integral equation for $h(u)$ 
that now includes a coupling to the higher-channel modes 
\[
h_n(u)=\left[-i f_n\left(k=2\sqrt{\Omega_B}\
 u\right)/\bar{k}\right]_{\bar{k}\to 0}.
\]
The real and imaginary parts of $\{h(u),h_n(u)\}$ 
decouple in the resulting equations.
For clarity, we show only the real part, which is
sufficient to analyse the effect of the higher channels
on the scattering length $a_{ad}$.
After some algebra, we find 
\begin{eqnarray}\label{sytem1}
&& \int_{-\infty}^{\infty}  \frac{d u'}{2 \pi u'^2}
 [ G^{0,0}(u,u') h(u') -  G^{0,0}(u,0) h(0)] \\ \nonumber
&& - \frac{\sqrt{\Omega_B}}{2u^2} \mathcal{L}\left(
\Omega_B[1+3u^2/4]\right )  h( u)
 = \frac{a_{ad}\sqrt{\Omega_B} }{a_\perp} G^{0,0}(u,0) \\ \nonumber
&& 
+ \Omega_B \sum_{n'\ne 0} \int_{-\infty}^{\infty} \frac{d u'}{\pi}
G^{0,n'}(u,u') h_{n'} (u'),
\end{eqnarray}
where the matrix elements $G^{n,n'}(u,u')$ can be extracted
from the matrix elements $A^{n,n'}_{k,k'}$
defined in Eq.~(\ref{lomega}), with $G^{n,n'}(u,u') = \Omega_B
A^{n,n'}_{k,k'}$ and the rescaling $k=2\sqrt{\Omega_B} u$.
We shall specify them in limiting cases below, but their
general form for arbitrary parameters is of no interest here.
Note that $G^{0,0}(u,u')=G(u,u')$ is given by Eq.~(\ref{Guu}). 
Furthermore, only modes with $m=m'=0$ are
coupled to the lowest transverse mode
and thus have to be kept.  Clearly,
Eq.~(\ref{inteqfin}) is reproduced, but
now includes a correction due to the higher channels.
The integral equation is then closed by 
\begin{equation}\label{sytem2}
\begin{split}
& \Omega_B^{-3/2}
  \int_{-\infty}^{\infty} \frac{d u'}{2 \pi u'^2}
 [ G^{n,0}(u,u')h(u') -  G^{n,0}(u,0) h(0)] \\[2mm]
&+ \mathcal{L} \left(n+\Omega_B[1+3u^2/4]\right)  h_n (u)
= \frac{a_{ad}}{ a_\perp} \frac{G^{n,0}(u,0)}{\Omega_B} \\[2mm]
&+ \Omega_B^{-1/2}
 \sum_{n'\ne0} \int_{-\infty}^{\infty} \frac{d u'}{ \pi}
G^{n,n'}(u,u')  h_{n'} (u').
\end{split}
\end{equation}
The system of integral equations given by Eqs.~(\ref{sytem1})
and (\ref{sytem2}) will now be analyzed in the two limiting cases.
In fact, we will see that in the BCS limit $a_\perp/a\to -\infty$,
higher channels are completely negligible, while
in the opposite dimer limit, they cause a renormalization
of $a_{ad}$ but no {\sl qualitative} change in the picture put
forward in the last section.  Moreover, in the dimer limit, 
we can solve the problem analytically and establish a connection
to the solution of the unconfined (3D)  problem \cite{skorniakov}.  
Since the effect of higher channels does not cause profound changes
even in the dimer limit, we conclude that
 the physical picture of Sec.~\ref{sectrans}
is reliable and qualitatively correct for all $a_\perp/a$.

\subsection{BCS limit}
\label{highbcs}

We now show that higher channels are indeed negligible
in the BCS limit, $\Omega_B\ll 1$. 
As the channel index $n$ has to be compared with the reduced
energy $\Omega_B$, it is intuitively clear that only small values
of $n$ can contribute.
In addition, relevant wavevectors obey $k,k'\propto \sqrt{\Omega_B}$. 
To make this more quantitative, we specify $A^{n,n'}_{k,k'}$ 
in Eq.~(\ref{lomega}) for the BCS limit, 
which in turn determines $G^{n,n'}(u,u')$
appearing in Eqs.~\eqref{sytem1} and \eqref{sytem2}.
For $\Omega_B\ll 1$, the integral 
representation
\begin{eqnarray*}
A^{n,n'}_{k,k'} & \simeq & \int_0^{\infty} d t 
e^{-\{\Omega_B+n'+[k^{\prime
2}+ k^2+ kk^\prime]/4 \}  t } \\ &\times&
\int_0^{\infty} d x
e^{-x} L_n (x) L_{n'}(x/4)
\end{eqnarray*}
follows from Eq.~(\ref{lomega}).
Both integrals can be directly computed, 
and we find $A^{n,n'}_{k,k'}=0$ for $n>n'$ within these
approximations, otherwise there are small corrections of order unity.
For $n\leq n'$, with an $\Omega_B$-independent constant $C_{n n'}$,
we obtain
\[
A^{n,n'}_{k,k'} = \frac{C_{nn'}}{\Omega_B + (k^2+k'^2+k k')/4 + n'}. 
\]
Therefore, except for the open channel $n=n'=0$, 
$A^{n,n^\prime}_{k,k'}$ is always of order unity.
Let us then analyze the scaling of the various terms in Eq.~\eqref{sytem2}
as a function of $\Omega_B\ll 1$. 
On the left hand side, the first term scales as  $\Omega_B^{-1/2}$, and
the second as $h_n$. On the right hand side, the first
term is at most of order $a_{ad}/a_\perp \propto \Omega_B^{-1/2}$,
and the last term scales as $\Omega_B^{1/2} h_n$. Power
counting then gives $h_n \propto \Omega_B^{-1/2}$.
{}From Eq.~\eqref{sytem1}, we now see
that the last term (describing the effect of higher
channels)
scales as $\Omega_B^{3/2}$, and is thus negligible compared to
the leading  terms, which are of order unity.
This fact allows us to safely conclude that higher
transverse levels do not affect the low-energy scaling
behavior of $a_{ad}/a_\perp$ in the BCS limit.
This conclusion also holds for $b_{ad}$ as one can show using
a similar power counting reasoning.

\subsection{Dimer limit}
\label{highdimer}

\begin{figure}[t!]
\begin{center}
\includegraphics[scale=0.3]{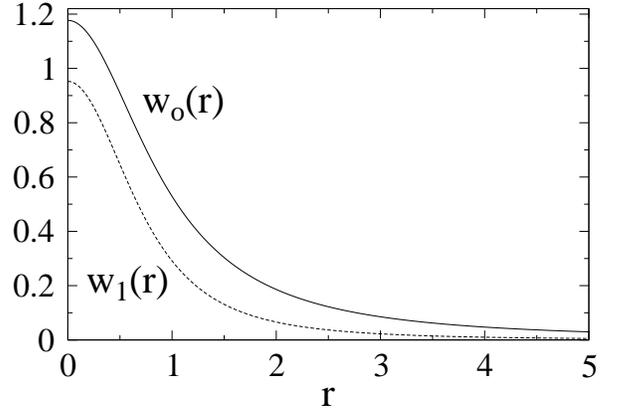}
\caption{Solutions $w_o (r)$ to Eq.~\eqref{free}, and $w_1 (r)$ to 
Eq.~\eqref{free2}.
\label{fig4} }
\end{center}
\end{figure}

The situation is quite different in the dimer limit, 
$a_\perp/a\to +\infty$, where excited
levels contribute to the asymptotic behavior of $a_{ad}$ and $b_{ad}$.
We first focus on the calculation of this correction
for $a_{ad}$. 
In the dimer limit, relevant values
for the channel number $n$ are of order $\Omega_B$, and
we later introduce the rescaled continuous variable $p=4n/3\Omega_B$
and convert the $n$-summation into an integration. 
Keeping only leading terms in $1/\Omega_B$,
Eqs.~\eqref{sytem1} and \eqref{sytem2} read
\begin{eqnarray}
\label{sytem21}
&& \frac 3 4 \frac{h(u)}{1+\sqrt{1+ 3 u^2/4}} =
\frac{a_{ad}}{a_\perp \sqrt{\Omega_B}} G^{0,0}(u,0)\\
\nonumber && + \frac 3 2  \int_{0}^{\infty} d p' \int_{-\infty}^{\infty}
\frac{d u'}{2 \pi} G^{0,p'}(u,u') \bar{h}_{p'} (u') 
\end{eqnarray}
and
\begin{eqnarray}\label{sytem22}
&&-2 \left( \sqrt{1+ 3 (u^2+p)/4 }-1 \right)  \bar{h}_p (u)
\\ && \nonumber
=  \frac{a_{ad}}{a_\perp \sqrt{\Omega_B}}  G^{p,0}(u,0)\\
&& \nonumber
+ \frac 3 2 \int_{0}^{\infty} d p' \int_{-\infty}^{\infty}
\frac{d u'}{2 \pi} \, G^{p,p'}(u,u') \bar{h}_{p'} (u'),
\end{eqnarray}
where we use $\bar{h}_p (u) = \Omega_B h_p (u)$. 
The second equation of this system is now independent
of the first one and can be solved consistently.
The kernel $G^{p,p'}(u,u')$ in the dimer
limit $\Omega_B\gg 1$ follows from $A_{k,k'}^{n,n^\prime}$,
 see Eq.~(\ref{lomega}),
which has the integral representation
\begin{eqnarray}
\nonumber
A_{k,k'}^{n,n'} &=& \int_0^{\infty} \frac{d t}{t} \, e^{- 
\{\Omega_B+n'+[k^{\prime 2}+k^2+ kk^\prime]/4\} \frac{3t}{4}  
 } \\ \label{akk}
&\times& \int_0^{\infty} d x \, L_n (x) L_{n'} (x/4) e^{-x / t}.
\end{eqnarray}
To evaluate the $x$-integral,  let us 
analyze the integral
\begin{eqnarray*}
I&=&\int_0^{\infty} d x \, e^{-x/t}  \, L_n (\lambda x) L_{n^\prime} (\mu x)\\
&=& t\int_0^{\infty} d y \, e^{-y}  \, L_n (\lambda ty) L_{n^\prime}
 (\mu t y) .
\end{eqnarray*}
Now we define  $\tilde{\lambda}=\lambda n t$ and $\tilde{\mu}=
\mu n' t$, which
stay constant in the limit $n,n'\to \infty$, and
use an asymptotic property of the Laguerre polynomials,
$\lim_{n\to \infty} L_n(x/n) = J_0 (2\sqrt{x}),$
where $J_0$ is a Bessel function.
Therefore, the integral $I$ in the limit of large $n,n^\prime$ 
is given by
\begin{eqnarray*}
I&=& t\int_0^\infty dy e^{-y}J_0
\left(2\sqrt{\tilde{\lambda} y} \right)
J_0\left(2\sqrt{\tilde{\mu} y} \right) \\ &=& 2
t\int_0^\infty xdx e^{-x^2}J_0\left(2\sqrt{\tilde{\lambda}} \ x\right)
J_0\left(2\sqrt{\tilde{\mu}} \ x \right).
\\ &=& te^{-(\tilde{\lambda}+\tilde{\mu})} I_0\left( 2
\sqrt{\tilde{\lambda}\tilde{\mu}}\right).
\end{eqnarray*}
Setting $\lambda=1$ and $\mu=1/4$, using the integral
representation for the Bessel function $I_0$,
\[
I_0(x)=\int_0^{2\pi}\frac{d\varphi}{2\pi}e^{-x\cos\varphi},
\] 
and performing the $t$-integral in Eq.~(\ref{akk}), we 
obtain
\begin{eqnarray*}
A_{k,k'}^{n,n^\prime} &=& \frac 4 3 \,\int_0^{2 \pi} \frac{d \varphi}{2 \pi}
\, \Bigl[ \Omega_B + (k^2+k'^2+k k')/4 \\ \nonumber
& +& 4( n + n^\prime + \sqrt{nn^\prime} \cos \varphi)/3 \Bigr]^{-1}.
\end{eqnarray*}
After rescaling $k =2 \sqrt{\Omega_B} u$ and
$n = 3 p /4\Omega_B$, we finally 
arrive at the expression
\begin{eqnarray*}
G^{p,p'}(u,u') &=& \frac 4 3 \,\int_0^{2 \pi} \frac{d \varphi}{2 \pi}
\, \Bigl[ 1 + u^2+u'^2+ u u'\\
& +& p + p' + \sqrt{pp'}
\cos \varphi \Bigr]^{-1}.
\end{eqnarray*}
Inserting this result into Eq.~\eqref{sytem22},
we can identify $(u,\sqrt{p},\varphi)$ as the
 cylindrical coordinates of a 3D vector
${\bf r}$ (which is of course still a momentum operator), where
$\bar{h} ({\bf r}) $ is now a function of ${\bf r}$. Writing
$h_o ({\bf r}) = r^2 \bar{h} ({\bf r})$, we obtain
for the integral equation of the higher-channel modes \eqref{sytem22}
the form
\begin{eqnarray}  \nonumber
&& \frac{\sqrt{1+ 3 r^2/4}-1}{r^2} \, h_o ({\bf r}) + \int
\frac{d {\bf r'}}{2 \pi^2} \frac{h_o( {\bf r'} )}{r'^2 (1+r^2+r'^2+
{\bf r} \cdot {\bf r'})} \\ 
\label{skor1}
 && = -\frac 2 3 \,
\left(\frac{a_{ad}}{a_\perp \sqrt{\Omega_B}} \right) \, \frac{1}{1+r^2}.
\end{eqnarray}
This integral equation is exactly the one
governing the fermionic three-body problem 
without confinement \cite{skorniakov}.
The symmetry of Eq.~(\ref{skor1}) implies that $h_o ({\bf r})$
only depends on  $r=|{\bf r}|$. We therefore write
\begin{equation}\label{wr}
h_o ({\bf r}) = - (2/3) \left
(a_{ad}/a_\perp \sqrt{\Omega_B}\right) w_o (r),
\end{equation}
where $w_o (r)$ is the solution to 
\begin{eqnarray} \nonumber
& & \frac{1}{\pi} \int_0^{\infty} \frac{d r'}{r \, r'}
\ln \left( \frac{1+r^2+r'^2+ r\,r'}{1+r^2+r'^2- r\,r'} \right)
w_o (r') \\ 
\label{free}
&+& \frac 3 4 \frac{w_o (r)}{1+\sqrt{1+3 r^2/4}} = \frac{1}{1+r^2}.
\end{eqnarray}
The numerical solution to this equation is shown in Fig.~\ref{fig4}.
We find $w_o (0) \simeq 1.179$, in accordance
with STM's result $w_o(0)\approx 1.2$ \cite{skorniakov,petrov03}.

We now go back to Eq.~\eqref{sytem21} and consider $u=0$.
Straightforward algebra then gives to leading order in $1/\Omega_B$ the
atom-dimer scattering length in the dimer limit as
\begin{equation} \label{kappa}
-\frac{a_{ad}}{a_\perp  \sqrt{\Omega_B}} 
\equiv  \kappa_\infty = \frac{3}{4  w_o (0)} \simeq 0.636.
\end{equation}
This value should be compared to $\kappa_\infty=9/32 = 0.28125$,
which results when higher channels are neglected, 
see Eq.~(\ref{dimer1}).  Note that our result makes 
explicit contact to the previous solution for the unconfined
case \cite{petrov03,skorniakov}. This is also seen by 
writing Eq.~(\ref{kappa}) as
\begin{equation}\label{connec-two}
a_{ad} = - \frac{a_{red,\perp}^2} { 2  (w_o(0) \, a) } 
\end{equation}
with the confinement scale $a_{red,\perp}=
(3 \hbar /2 m_0 \omega_\perp)^{1/2}$ for reduced mass 
$2 m_0/3$ of the atom-dimer complex.
  With the 3D atom-dimer scattering length $w_o(0) a=w_o(0)/\kappa_B$, 
this result exactly matches the 
dimer limit of the analogous two-body result (\ref{a1d}).
Equation (\ref{connec-two}) predicts that the product of
the 1D and 3D atom-dimer scattering lengths (in units of $a_\perp$)
is a universal constant, independent of the atomic properties.
In fact, we shall see later that it is even independent of statistics.

Let us then turn to the determination of $b_{ad}$ in the dimer limit,
including the contribution of higher transverse channels. 
The calculation follows closely the one for $a_{ad}$,
and  we shall therefore only briefly outline the various steps. 
Starting with the imaginary part of the general coupled equations
resulting from \eqref{integral-equa}, we arrive at 
new coupled equations very similar to Eqs.~\eqref{sytem1}
and \eqref{sytem2}. We then use the
asymptotic expression for $A_{k,k'}^{n,n^\prime}$ in the dimer limit and
perform the rescaling $k =2 \sqrt{\Omega_B} u$,
$n = 3 p /4\Omega_B$ and $\bar{h}_p (u) = \Omega_B h_p (u)$. 
Once again we identify 
$(u,\sqrt{p},\varphi)$ as  the  cylindrical coordinates of a 3D vector
${\bf r}$, and, in order to obtain an integral equation with
a spherically symmetric solution, we define ${\rm Im} \, \bar{h} ( {\bf r})
= - (u/r^2)\,(1/\Omega_B)\, w_1 (r)$. After some algebra, we eventually
obtain an integral equation for $w_1 (r)$ very similar to 
Eq.~\eqref{free}, namely
\begin{equation}\label{free2}
\frac{3}{4} \frac{w_1 (r)}{1+\sqrt{1+3 r^2/4}} + \frac{1}{\pi}
\int_0^\infty d r' K(r,r') w_1 (r') = \frac{1}{3 (1+r^2)^2},
\end{equation}
with the kernel
\begin{equation}\label{kernelb}
K(r,r') = \frac{1}{r^2} \left(2 - \frac{1+r^2+r'^2}{r r'}
\ln \left( \frac{1+r^2+r'^2+ r r'}{1+r^2+r'^2- r r'} \right) \right).
\end{equation}
The numerical solution to this equation is shown in Fig.~\ref{fig4}.
Including  the higher channels, we thus find for $b_{ad}$ instead
of Eq.~(\ref{bad}) the result
\begin{equation}
\frac{b_{ad}}{a_\perp} = \frac{w_1 (0)}{\Omega_B^{3/2}}, \quad w_1 (0) \simeq 0.952.
\end{equation}
Notably, this is essentially the same behavior as in 
Eq.~\eqref{bad}, where the higher channels were neglected.
Even the coefficient of the asymptotic behavior, $0.952$ compared
to  $8/9 \simeq 0.889$, is very close.

To summarize, we conclude that
in the deep dimer limit, the relevant physics is not 
changed by the confinement potential.

\section{Bosonic three-body problem}\label{secbos}

In this section, we consider the problem of three identical bosons 
in a tight transverse harmonic confinement, allowing to analyze the 
1D bosonic analogue of the problem studied up to now for fermions. 
The main difference is that the three-body wavefunction
is now totally symmetric. On a formal level, the
calculation is similar to the fermionic one,
so in what follows we go through it, highlighting the
differences and keeping the same notation as much as possible.
The definition of the coordinate system remains the same,
i.e., ${\bf x_1}$, ${\bf x_2}$ and  ${\bf x_3}$ are the 
3D positions of the bosons.
Next the orthogonal transformation (\ref{orthogonal})
to variables $({\bf x}, {\bf y},{\bf z})$ is performed.
The harmonic trap potential is still
diagonal in the positions after this transformation, and
the center-of-mass coordinate ${\bf z}$ again decouples.
Hence the three-body problem reduces to 
\begin{eqnarray}\label{schrob}
&~&\left( - \frac{\hbar^2}{m_0}\nabla_{\bf X}^2 + U_c({\bf X}) -E\right)
 \Psi({\bf X}) \\&~& = -
 \left[V({\bf y})+ \sum_{\pm}  V ( {\bf r}_{\pm} )  
 \right]\Psi({\bf X}).\nonumber
\end{eqnarray}
As in Eq.~(\ref{rpm}), ${\bf r}_{\pm}$ denotes
the distance between the boson at ${\bf x_1}$  and the
${\bf x_2}$ or ${\bf x_3}$ particles, respectively, while 
${\bf y}={\bf x_2}-{\bf x_3}$.
Note that all three bosons interact, leading to the 
extra term $V({\bf y})$ in Eq.~\eqref{schrob}
compared to the fermionic
Eq.~\eqref{schro}. 
Incorporating atom-atom interactions via
the pseudopotential approach, 
one boundary condition now reads 
\begin{equation}\label{bc1}
\Psi ({\bf X})\simeq  \frac{f({\bf x})}{4 \pi} \left[
\frac{1}{y} - \frac{1}{a} \right], \quad
{\bf y} \to 0,
\end{equation}
while the other two are 
\begin{equation}\label{bc2}
\Psi ({\bf X})\simeq \frac{f({\bf r}_{\perp,\pm})}{4 \pi} \left[
\frac{1}{ r_{\pm}} - \frac{1}{a} \right], \quad
{\bf r}_{\pm} \to 0,
\end{equation}
where ${\bf r}_{\perp, \pm} = -\cos(\theta){\bf x} \pm 
\sin(\theta) {\bf y}$ 
carries an extra minus sign compared to Eq.~(\ref{ortho1}).
Because the three-body wavefunction is fully symmetric, the
conditions (\ref{bc2}) are redundant, and it is sufficient
to satisfy only Eq.~(\ref{bc1}) in what follows.
As discussed in the Introduction, the bosonic three-body problem
requires a short-distance regularization $R^*$, which we implement
using Petrov's scheme \cite{petrovreg}.
The regularization (\ref{Rstar}) is formulated via the action of a 
differential operator on the scattering
amplitude that describes the relative kinetic and confinement energy.
The boundary condition \eqref{bc1} (for ${\bf y}\to 0$) is thereby modified to
\begin{equation}\label{bc1b}
\Psi({\bf X}) \simeq  \left[
\frac{1}{y} - \frac{1}{a}-R^*\left(\frac{m_0 E}{\hbar^2}  +\nabla^2_{{\bf x}}-
\frac{2 \rho_{\bf x}^2}{a_\perp^4}\right) \right] \frac{f({\bf x})}{4 \pi}, 
\end{equation}
where $\rho_{\bf x}$ is the component of ${\bf x}$ along the transverse
direction.
The boundary condition for the two-body
problem also changes, now leading to a modified equation
for the dimensionless binding energy $\Omega_B$ defined
in Eq.~(\ref{omegab}). Instead of 
Eq.~({\ref{bs}), it now reads
\begin{equation}\label{dimermod}
\zeta(1/2,\Omega_B)+a_\perp/a-4R^*\Omega_B/a_\perp=0.
\end{equation}
In the dimer limit, $\Omega_B\gg 1$,
the 3D result \cite{petrovreg} follows,
\begin{equation}\label{kpdimer}
\kappa_B=\frac{1}{2R^*}\left(\sqrt{1+\frac{4R^*}{a}}-1\right),
\end{equation}
while in the BCS limit, $R^*$ gives only subleading corrections. 

In an identical way as for fermions, we then obtain the 
bosonic analogue of Eq.~\eqref{final},
take the ${\bf y} \to 0$ limit 
according to Eq.~\eqref{bc1b},  and finally
obtain the integral equation like
in Eq.~\eqref{integral-equa}.  This modified integral equation 
for bosons reads 
\begin{equation}\label{inteqbos}
\widetilde{{\cal L}}(\Omega_\lambda) f_\lambda  = 
\sum_{\lambda'} \widetilde{A}_{\lambda,\lambda'} \, f_{\lambda'},
\end{equation}
where $\Omega_\lambda$ is given by Eq.~(\ref{omegalam}),
and the matrix elements are
\begin{equation}\label{Astar}
\begin{split}
\widetilde{A}_{\lambda,\lambda'}  = -\frac{8\pi \hbar^2 \,a_\perp}{m_0}
\int d {\bf x} & \, \psi_{\lambda}^* ( {\bf x})\,
\psi_{\lambda'} ( - {\bf x} \, \cos  \theta) \, \times \\[2mm]
& \times G_{E-E_\lambda'} (  {\bf x} \, \sin \theta,0 ). \end{split}
\end{equation}
When compared to the fermionic Eq.~(\ref{lomega}), the bosonic 
Eq.~\eqref{Astar} carries an overall factor of $-2$, implying that
an effective repulsion has turned into an attractive force.
Moreover, the integrand in 
Eq.~(\ref{Astar}) remains unaltered since $\cos 2 \theta = -\cos
\theta$ and $\sin 2 \theta = \sin \theta$.
  In addition, the ${\cal L}$ function (\ref{lome}) is modified to
\begin{equation} \label{lmodified}
\widetilde{\cal L}(\Omega_{\lambda}) =  
\zeta(1/2, \Omega_{\lambda}) - \zeta(1/2,\Omega_B) - R^* (k^2-\bar{k}^2+2n),
\end{equation} 
where from now on, all lengths (momenta) will again be given
in units of $a_\perp$ $(1/a_\perp)$. 
Note that the distance between
the center-of-mass of the two bosons ${\bf x_{2,3}}$
and the one at ${\bf x_1}$ is ${\bf x} \sin\theta$. Since
$\sin\theta=\sin(2\theta)$,
exactly the same rescaling as in the fermionic case will be
employed in what follows.

\subsection{Scattering solution}\label{scabosons}

Let us then proceed by
projecting Eq.~\eqref{inteqbos} onto the ground state. 
We use the same scattering
Ansatz (\ref{skor}) as for fermions, and after expanding in 
$\bar{k}=2 \sqrt{\Omega_B} \ \bar{u}$, one easily
obtains the bosonic version of Eq.~\eqref{inteqfin},
\begin{eqnarray} 
&& \int_{-\infty}^{\infty} \frac{d u'}{ 2\pi u'^2} 
[ G(u,u') h(u') -  G(u,0) h(0)]  
 \label{inteqfinbos} \\ \nonumber
&& + \frac{ \sqrt{\Omega_B}}{4u^2} 
\widetilde{{\cal L}}(u)   h(u)
 = \frac{a_{ad}\sqrt{\Omega_B}}{a_\perp} G(u,0) +iH(u),
\end{eqnarray}
with the functions $G(u,u')$ and $H(u)$ defined 
as for fermions, see Eqs.~(\ref{Guu}) and (\ref{hu}).
Furthermore, Eq.~\eqref{lmodified} gives 
\begin{eqnarray} \label{lrescaled}
\widetilde{\cal L}(u) & = &  \zeta\left(1/2, 
\Omega_B(1+3u^2/4)\right)\\ 
&-& \zeta(1/2,\Omega_B)
-3 R^* \Omega_B   u^2. \nonumber
\end{eqnarray} 
In the bosonic case, the atom-dimer scattering length 
$a_{ad}/a_\perp$ therefore  depends on
the two dimensionless parameters $\Omega_B$ and $R^*/a_\perp$. In that sense,
the bosonic problem is non-universal \cite{hammer}.

Let us start with the {\sl dimer limit}, $\Omega_B\gg 1$.
In that case
 it is useful to introduce the dimensionless regularization parameter
\begin{equation}\label{rstar}
r^*=\kappa_B R^* = \frac{1}{2}\left(\sqrt{1+\frac{4R^*}{a}}-1\right).
\end{equation}
The solution to Eq.~\eqref{inteqfinbos} in the dimer limit can be 
found analytically again, with the result
\begin{equation}\label{asympboson}
\frac{a_{ad}}{a_\perp }
= \frac{9}{64 (1+2 r^*)} \sqrt{\Omega_B} 
+ \frac{\beta}{\sqrt{\Omega_B}},
\end{equation}
where the coefficient $\beta$ is given for $r^*=0$ as
\[
\beta=9/256 + (3 \sqrt{3} + 4 \pi)/8 \pi +3/64
\simeq 0.6638.
\]
 We also find $b_{ad} / a_\perp = - (4/9)\Omega_B^{-3/2}$, which 
resembles the fermionic equivalent \eqref{bad}.

However, as in the fermionic case,
Eq.~\eqref{inteqfinbos} is not sufficient in the dimer limit,
 and higher transverse
channels must be included.
Skipping details of the calculation -- which closely parallels 
the fermionic one in Sec.~\ref{highdimer} -- 
and keeping the same notation, we find instead of Eq.~\eqref{skor1}  
the bosonic scattering solution
\begin{eqnarray}\label{h_o}
&&\frac{\sqrt{1+ 3 r^2/4}-1+3r^*r^2/4}{r^2} \, h_o ({\bf r}) \\
&&- \int \frac{d {\bf r'}}{ \pi^2} \frac{h_o( {\bf r'} )}{r'^2 (1+r^2+r'^2+
{\bf r} \cdot {\bf r'})} =\nonumber\\&& +\frac{4}{3} \, 
\left(\frac{a_{ad}}{a_\perp \sqrt{\Omega_B}} \right) \, \frac{1}{1+r^2}.
\nonumber
\end{eqnarray}
Defining the function $w_o(r)$ as in Eq.~\eqref{wr},
we obtain after angle integration instead of Eq.~(\ref{free}) the equation
\begin{eqnarray}\nonumber
&&\frac{\sqrt{1+ 3 r^2/4}-1+3r^*r^2/4}{2r^2} \, w_o  (r) =
-\frac{1}{1+r^2}\\
&+&\frac{1}{\pi} \int_0^{\infty} \frac{d r'}{r \, r'}
\ln \left( \frac{1+r^2+r'^2+ r\,r'}{1+r^2+r'^2- r\,r'} \right)
w_o (r').
\label{skorniakovbos}
\end{eqnarray}
At $r^*=0$, this is precisely STM's Eq.~(15) \cite{skorniakov},
once the 3D scattering length is identified with $w_o(0) a_B=w_o(0)/\kappa_B$.
For finite $r^*$, it is equivalent to the equation recently
studied by Petrov, since Eq.~\eqref{skorniakovbos}
follows from Eq.~(12) of Ref.~\cite{petrovreg} upon substitution of the
standard 3D scattering Ansatz.
The scattering length can be extracted in a way similar to the fermionic
case. For that purpose, we put $u=0$ in the bosonic version of Eq.~\eqref{sytem21}, 
and arrive at
\[
-\frac{3}{8} (1 + 2 r^*)=-\frac{8}{3}\frac{a_{ad}}{a_\perp \sqrt{\Omega_B}}
\left[1-\frac{2}{\pi}\int_0^\infty dr\frac{ w_o (r)}{1+r^2}\right].
\]
On the other hand, taking $r\to 0$ in Eq.~\eqref{skorniakovbos} yields
\[
\frac{3}{8}(1 + 2 r^*)\frac{w_o(0)}{2}=
-1+\frac{2}{\pi}\int_0^\infty  dr \frac{ w_o(r)}{1+r^2}.
\]
Consequently we have
\begin{equation}\label{aadb}
\frac{a_{ad}}{a_\perp \sqrt{\Omega_B}}=-\frac{3}{4w_o(0)},
\end{equation}
i.e., the same relation as for fermions, leading to Eq.~\eqref{connec-two}
and matching the analogous two-body result \eqref{a1d}.
We conclude that in the dimer limit, for both fermions and bosons,
 the 1D scattering length $a_{ad}$ is 
always inversely proportional to the 3D scattering length
$w_o(0)/\kappa_B$. The latter has been studied in
detail in Ref.~\cite{petrovreg} as a function of $R^*/a$.
It was found to diverge whenever a new Efimov state
splits from the continuum, going through zero in between 
the resonances, see Fig.~1 of Ref.~\cite{petrovreg}. According
to Eq.~(\ref{aadb}), the atom-dimer scattering length $a_{ad}$
will behave in the opposite manner, i.e., it vanishes every time
a new bound state emerges and diverges in between.
Note that in order to derive Eq.~\eqref{h_o}, 
one needs to impose $w_o(0)/\kappa_B<a_\perp$, 
which follows from $\Omega_B\gg 1$, so that 
divergences of $w_o(0)/\kappa_B$  will in practice be
smeared out on lengthscales $\approx a_\perp \gg a$.

In the limit of large $R^*$, one can solve Eq.~\eqref{skorniakovbos}
analytically by expanding in inverse powers of $r^*$ \cite{petrovreg},
\[
w_o(r)=-\frac{8}{3r^*}\frac{1}{1+r^2}+{\cal O}(1/r^{*  2}).
\]
Taking into account Eqs.~(\ref{kpdimer}) and (\ref{rstar}), it 
is easily seen that the 3D scattering length becomes independent of $R^*$,
while for the 1D scattering length, the contribution of higher
channels clearly becomes negligible. 
We then obtain from Eq.~\eqref{asympboson}
\begin{equation}\label{aaddimerbos}
a_{ad}=\frac{9}{64}\frac{a_\perp^2}{a},
\end{equation}
which is a universal result in the sense that it does not
depend on $R^*$.

Higher channels can also be taken into account exactly 
for the calculation of $b_{ad}$ in the dimer limit.
We follow the fermionic case in Sec. \ref{highdimer} and obtain
\begin{equation}
\frac{b_{ad}}{a_\perp} = - \frac{w_1 (0)}{\Omega_B^{3/2}},
\end{equation}
where, in contrast to the fermionic case, $w_1(r)$ solves
the integral equation
\begin{equation}\label{eqw1}
\begin{split}
\frac{3}{4} & \frac{w_1 (r)}{1+\sqrt{1+3 r^2/4}} + \frac{3}{8} 
r^* w_1(r)  \\[2mm]
&-  \int_0^\infty \frac{d r'}{\pi} K(r,r') w_1 (r') = \frac{1}{3 (1+r^2)^2},
\end{split}
\end{equation}
with $K(r,r')$ given by Eq.~\eqref{kernelb}. The solution for $w_1(0)$ 
as a function of $R^*/a$ is shown in Fig.~\ref{w10}. 
\begin{figure}[t!]
\begin{center}
\includegraphics[scale=0.30]{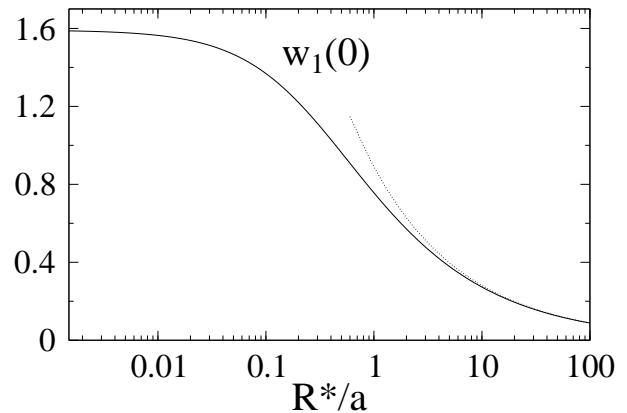}
\caption{
Solution $w_1(0)$ (full line) to Eq.~\eqref{eqw1} as a function of $R^*/a$.
The dotted curve gives the
asymptotic behavior $w_1(0)= (8/9)\sqrt{a/R^*}$ for large $R^*/a$.
  \label{w10} }
\end{center}
\end{figure}
Apparently, there
is no divergence due to Efimov states. In particular, the $R^* \to0$
limit is well-defined, with $w_1(0) \simeq 1.59$, and
\begin{equation}
b_{ad} \simeq - 12.7 a^3/a_\perp^2,
\end{equation}
validating a zero-range atom-dimer potential in the low-energy limit.
The other limit is $R^* \gg a$, where $w_1(0) \simeq (8/9)
\sqrt{a/R^*}$, and therefore
\begin{equation}
b_{ad} = - (64/9) R^* ( a/a_\perp)^2.
\end{equation}
Comparing this result to Eq.~\eqref{aaddimerbos}, we find that the zero-range
 potential approximation requires $R^*a^3/a_\perp^4 \ll 1$, which
self-consistently holds for $\Omega_B \gg 1$.

In the BCS limit, $\Omega_B\ll 1$,
the same argument as in Sec.~\ref{highbcs} 
can be applied to the bosonic case.  Therefore higher channels
can again be disregarded in that limit. 
Moreover, the regularization parameter $R^*$ drops out to leading 
order, since relevant momenta are small
($\propto \sqrt{\Omega_B}$) and the short-ranged cutoff ceases to matter.
In this sense, the three-body problem becomes universal again in the
BCS limit.
The asymptotic behavior of $a_{ad}$ and $b_{ad}$ can be analytically computed
using the Bethe Ansatz, see Sec.~\ref{bethe}. 
One can verify that $h(u) = -2 u/(1+u^2)^2$
solves the imaginary part of Eq.~\eqref{inteqfinbos} in the BCS limit,
leading to 
\begin{equation}\label{bsol}
\frac{b_{ad}}{a_\perp} = \frac{h'(0)}{2 \sqrt{\Omega_B}} = - \, 
\frac{1}{\sqrt{\Omega_B}}.
\end{equation}
For the real part of Eq.~\eqref{inteqfinbos}, we obtain
from the Bethe Ansatz the solution  
\begin{equation} \label{realsol}
h(u) = 4 \, \frac{a_{ad} \sqrt{\Omega_B}}{a_\perp} \, \frac{u^2}{1+u^2}
\end{equation}
in the BCS limit.  Evidently, the condition
$h(0)=-1$ cannot be fulfilled for any finite $a_{ad}$.
Similarly to $b_{ad}$ for the fermionic case, 
the asymptotic behavior of $a_{ad}$ is therefore expected to  
diverge as $a_{ad} \propto - 1 /\Omega_B^{3/2}$ when
approaching the BCS limit.  We shall discuss this point in 
detail in Sec.~\ref{bethe}.

For the general case of arbitrary $a_\perp/a$, we have solved 
numerically  the real-space version of  Eq.~\eqref{inteqfinbos} as in 
Sec.~\ref{numerical}, neglecting higher channels.
For the sake of simplicity, we take $R^*=0$.
When compared to the fermionic Eq.~(\ref{realspace}),
the kernel (\ref{kdef0}) is now 
$K = K_1 - K_2/2$, with $K_{1,2}$ given 
in Eqs.~\eqref{kdef1} and \eqref{kdef2}, respectively.  The results
are shown in Fig.~\ref{aadbad}. Clearly, they closely match the predicted
behavior in both limits.
\begin{figure}[t!]
\begin{center}
\includegraphics[scale=0.30]{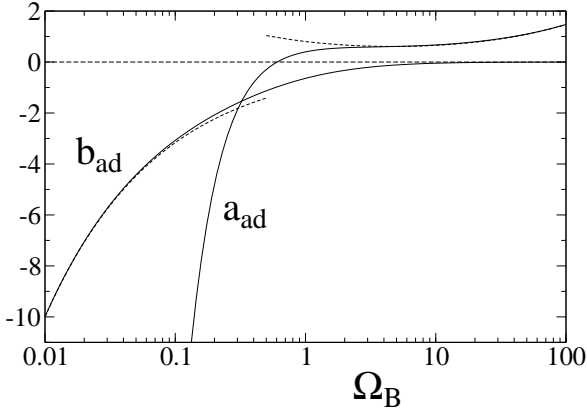}
\caption{
Plot of $a_{ad}$ and $b_{ad}$ (full lines) as function of
the dimensionless binding energy $\Omega_B$
for the three-boson problem at $R^*=0$.
The asymptotic behavior is shown (i) for $b_{ad}$ in the BCS
limit, $-1/\sqrt{\Omega_B}$, and  (ii) for $a_{ad}$ in the dimer limit,
$(9/64) \sqrt{\Omega_B}+ 0.6638/\sqrt{\Omega_B}$.  \label{aadbad} }
\end{center}
\end{figure}
A finite $R^*$ in Eq.~(\ref{realspace}) gives similar results.
{}From the above discussion, this is clear in the
BCS limit.  Moreover, when just keeping the lowest
transverse channel, the qualitative behavior is not 
modified in the dimer limit either.
Therefore we expect a very similar picture for arbitrary $R^*$ 
as the one shown for $R^*=0$ in Fig.~\ref{aadbad}, as long
as higher channels can be neglected [as presumed 
in Eq.~(\ref{realspace})]. While this approximation is justified in the 
BCS limit, in the dimer limit it does not 
capture the correct asymptotic behavior, except for large $R^*/a$,
see above.
In particular, the influence of Efimov states on $a_{ad}$ becomes
important and affects the physics in the dimer limit
in a significant way.
This is in contrast to the fermionic case, where the projection
onto the lowest transverse state already yields the qualitatively correct
behavior even in the dimer limit.

\subsection{Bound States: Trimers}\label{trimers}

For the bosonic problem, one expects atom-dimer bound states (trimers)
to form under certain conditions. We therefore next derive the
relevant integral equation. In this case, the scattering Ansatz
cannot be used, and the resulting homogeneous integral equation 
must be considered.
The total energy of the system is now $E=-2\, \hbar \omega_\perp\Omega_B-E_T$,
where the trimer binding energy is written as
\begin{equation}\label{trimerbinding}
E_T=\frac{3}{4} \frac{\hbar^2}{m_0} \kappa_B^2 u_0^2,
\end{equation}
with $u_0$ being the inverse size of the trimer 
in units of $a_B=1/\kappa_B$. Projecting 
Eq.~\eqref{inteqbos} onto the transverse ground state, we find
\begin{eqnarray} 
&~&\Bigl[\zeta\left(\frac{1}{2},\Omega_B(1+\frac{3}{4}(u^2+u_0^2)\right)-
\zeta\left(\frac{1}{2},\Omega_B\right)\nonumber\\
&-& 3 \Omega_B R^*(u^2+u_0^2)
\Bigr]f(u)\nonumber\\
&=&
-\frac{2}{\pi\sqrt{\Omega_B}}\int_{-\infty}^{\infty}du'
G_{u_0}(u,u')f(u'),\label{iebound}
\end{eqnarray}
where the function $G_{u_0}(u,u')$ is defined by
\[
G_{u_0} (u,u')= \sum_{p=0}^\infty \frac{4^{-p}}{1+3u_0^2/4+u^2+u^{\prime 2}+ 
u u^\prime + p/\Omega_B}.
\]
Equation \eqref{iebound} is then an eigenvalue equation for $u_0$.

First, we study Eq.~\eqref{iebound} in the dimer limit.
In order to obtain a nontrivial solution, we need to consider 
$u_0 \ll 1$. Then $f(u)$ is dominated by its
small-$u$ behavior, and using Eq.~(\ref{rstar}), 
Eq.~\eqref{iebound} is simplified to give
\begin{equation}\label{dimsolbos}
f(u) = \frac{1}{u_0^2+u^2}  
\frac{32}{9\pi \Omega_B (1+2 r^*)} 
 \int_{-\infty}^{\infty} d u' f(u'),
\end{equation}
which implies $f(u) \propto 1/(u^2+u_0^2)$. Self-consistency
of this expression leads to 
\begin{equation}\label{asymptobound}
u_0 = \frac{32}{9 \Omega_B (1+2 r^*)} \;.
\end{equation}
Note that we have supposed here that $f(u)$ is an even function.
Taking instead an odd function, we get
\[
f(u) = \frac{u}{u_0^2+u^2}\; \frac{32}{9} 
\frac{1}{\pi \Omega_B (1+2 r^*)} \;
 \int_{-\infty}^{\infty} d u' u' f(u'),
\]
which does not allow for a non-trivial and self-consistent solution.
Therefore $f(u)$ cannot be odd in the dimer limit.
For large $r^*$, the trimer binding energy
becomes universal, and is given by
\begin{equation}
E_T=\frac{1024}{27}\frac{\hbar^2}{m_0} \frac{a^2}{a_\perp^4}.
\end{equation}
Note that the existence of this novel {\it confinement-induced
trimer state} (CIT) is consistent with  
 the positive 1D atom-dimer scattering 
length for bosons found in the dimer limit, see Eq.~(\ref{aaddimerbos}).

In the BCS limit, Eq.~\eqref{iebound} for the bound states
reduces to 
\begin{eqnarray}
&&\left[\frac{1}{\sqrt{1+3(u^2+u_0^2)/4}}-1\right]f(u)\nonumber\\
&&=-\frac{2}{\pi}\int_{-\infty}^\infty\frac{du'f(u')}{1+
3u_0^2/4+u^2+u'^2+uu'}\;.\label{ieBCS}
\end{eqnarray}
The fact that one can use the Bethe Ansatz to solve directly
the BCS limit also holds for bound states.
Indeed, we will verify in Sec.~\ref{bethe}
that Eq.~\eqref{ieBCS} follows from a 1D Schr\"{o}dinger equation,
see Eq.~\eqref{ieBCS2} below, and leads to the exact solution
\begin{equation}\label{BCSsolbos}
f(u)=\frac{1}{u^2+u_0^2}
\end{equation}
for $u_0=2$.
We have verified numerically that there are no other solutions,
neither even nor odd. In fact, in real space, Eq.~\eqref{BCSsolbos} 
is nothing but the known bound-state
Bethe function for three attractively interacting bosons \cite{gaudin},
\begin{equation}\label{BAtrimer}
\psi\propto \exp\left[-
(|z_1-z_2|+|z_2-z_3|+|z_3-z_1|)/a_{aa}\right].
\end{equation}

In between the BCS and the dimer limits, we have investigated
Eq.~\eqref{iebound} numerically, see Fig.~\ref{bound}.
Qualitatively, we see that the trimer energy barely depends 
on $R^*$, and in the BCS limit even becomes completely independent
of this regularization parameter.
However, Eq.~\eqref{iebound} is only approximate since it is derived 
by projecting Eq.~\eqref{inteqbos} onto the lowest transverse state,
and therefore does not include the effect of higher channels.
Notably, Efimov trimer states cannot be recovered within this 
approximation. Such states are irrelevant in the BCS limit, where the
lengthscale $a_{aa}$ of the confined trimer state [see Eq.~(\ref{BAtrimer})]
is larger than the typical lengthscale $a$ of Efimov states,
 so that the two problems decouple. However,
when moving towards the dimer limit, the interplay between the 
confined trimer state and Efimov states plays a role.
\begin{figure}[t!]
\begin{center}
\includegraphics[scale=0.30]{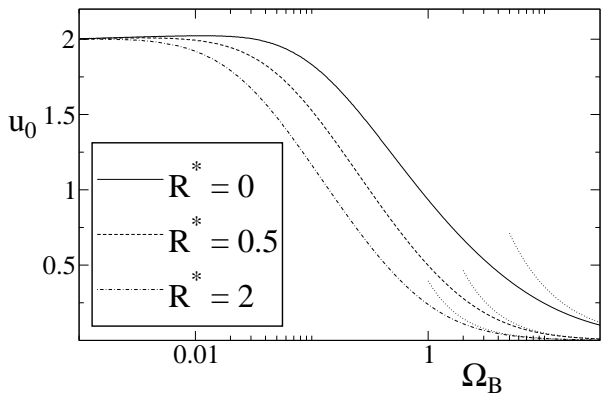}
\caption{Parameter $u_0$ appearing in the trimer binding 
energy, Eq.~\eqref{trimerbinding}, as a function
of $\Omega_B$ for various values of  $R^*/a_\perp$.
The dotted lines correspond to the asymptotic behavior in the dimer
limit, Eq.~\eqref{asymptobound}. 
\label{bound} }
\end{center}
\end{figure}

In the dimer limit, it is possible to study the CIT
in a quantitative manner, including the way
it is affected by Efimov trimer states. In fact, we have shown
in Sec.~\ref{scabosons}, for the dimer limit, that the low-energy 
atom-dimer scattering properties are described by a 1D contact 
potential $V(z) =  (-2 \hbar^2 /m_0 a_{ad})\delta(z)$. This potential
has a bound state for $a_{ad}>0$ and its energy is given by
\begin{equation}\label{bound1d}
E_T = \frac{\hbar^2}{m_0 \, a_{ad}^2}.
\end{equation}
The corresponding 1D scattering length $a_{ad}$ is
related through Eq.~\eqref{connec-two}
to the 3D atom-dimer scattering length 
$w_o(0)a_B$ (i.e., without external confinement)
obtained from the solution of Eq.~\eqref{skorniakovbos}.  This quantity is
shown, e.g., in Fig.~1 of Ref.~\cite{petrovreg}. 
The bound state of this 1D contact potential -- present 
only for $a_{ad} >0$ -- describes the CIT as long
as its energy $E_T < \hbar \omega_\perp$.
This implies $a_{ad} > a_\perp$, which in turn is equivalent to
the condition $w_o(0) a_B < a_\perp$.
This coincides with the  condition in Sec.~\ref{scabosons}
for the validity of Eq.~\eqref{h_o}, and therefore of Eq.~\eqref{connec-two}
for bosons.
The CIT is thus present only for
 $w_o(0) a_B<0$, and its energy follows from Eqs.~\eqref{bound1d} and
\eqref{connec-two} in the form
\begin{equation}
\frac{E_T}{\hbar \omega_\perp} =  \left( \frac{a}{a_\perp} \right)^2 \,
\frac{32}{9} \, \left( \frac{w_o(0) \, a_B}{a} \right)^2.
\end{equation}
{}From the numerical solution of Eq.~\eqref{skorniakovbos}, we obtain
results for $E_T$ as a function of $R^*/a$, see Fig.~\ref{energiet}.
\begin{figure}[t!]
\begin{center}
\includegraphics[scale=0.3]{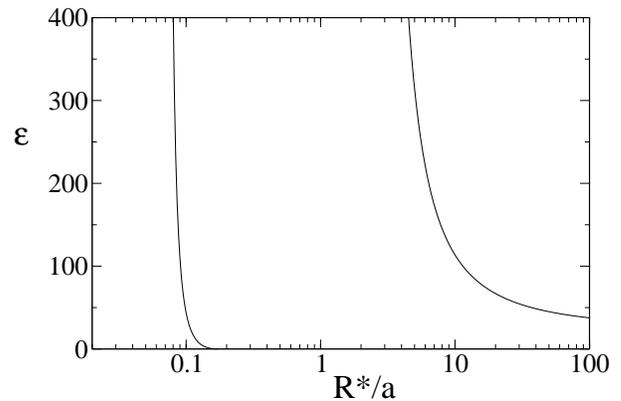}
\caption{ Reduced energy $\epsilon \equiv (E_T/\hbar \omega_\perp) 
(a_\perp/a)^2$ for the  CIT in the dimer limit
as a function of $R^*/a$.
\label{energiet} }
\end{center}
\end{figure}
Obviously, in the dimer limit,
the CIT only exists for certain values of $R^*/a$.
While it is present for all values larger than $R^*/a\simeq 2.2$,
below this threshold, it disappears and reappears periodically
for decreasing $R^*/a$.
One such cycle is shown in Fig.~\ref{energiet}, where 
the CIT has disappeared in the window $0.17\alt R^*/a \alt 2.2$,
but reappears for $R^*/a\alt 0.17$.
For $R^* \ll a$, the behaviour is exactly periodic as a function
of $\ln (R^*/a)$ and can be obtained analytically using the known
 result  \cite{hammer} 
\begin{equation}
\frac{w_o(0)a_B}{a} = 1.46 - 2.15 \tan \left[ 1.00624 \ln (a \Lambda^*)
+ 0.09 \right], 
\end{equation}
together with $\Lambda^* \simeq 6.6/R^*$ \cite{petrovreg}.
The ultimate fate of the CIT and its evolution in between
the BCS and the dimer limits
remains a difficult open question that is
outside the scope of our paper.

\section{Bethe Ansatz in the BCS limit}\label{bethe}

In the BCS limit, $\Omega_B \to0$, the role of transverse
excited states is negligible. Confined in the transverse ground
state, the motion of particles becomes effectively
1D, with contact $\delta$-interactions between bosons or
between distinguishable fermions. This strong simplification of
the original $3$D hamiltonian hence allows for exact solutions on the
many-body level, namely the Lieb-Liniger
model for bosons \cite{lieb}, or the Yang-Gaudin model for
fermions \cite{yg}, both based on the powerful Bethe Ansatz.
Recently, these two models have been used to study
many-body properties of cold atoms in
$1$D confined geometries \cite{tokatly,fuchs}.
In this section, we use the Bethe Ansatz \cite{gaudin}
 to solve the three-body problem
for bosons and fermions in the BCS limit. 
This leads to analytical predictions for $a_{ad}$ and $b_{ad}$,
 as well as for the bosonic trimer state.
 We thereby derive some of the results presented
 in previous sections. Noting that the atom-dimer scattering does
not lead to any reflected wave, we also clarify the connection
between the divergence of  $a_{ad}$ ($b_{ad}$) for bosons (fermions) 
and the fact that the scattering process is reflectionless.
The reduction of the original problem to a 1D model
is not completely straightforward and needs to be 
considered cautiously. Even in the BCS limit, the few-body
wavefunction is not restricted to the transverse ground state.
In order to build up a boundary condition like Eq.~\eqref{asymto}, 
imposed by the two-body
potential for small particle distances,
many excited transverse states have to be involved. However, as soon
as distances between particles are larger than $a_\perp$, essentially only the 
transverse ground state can be occupied. In our case, the relevant
lengthscale is the 1D scattering length $a_{aa} = a_\perp^2/2 
|a|\gg a_\perp$, which justifies the neglect of transverse
excited states. This simplification becomes incorrect
for problems involving lengthscales $\le a$,
e.g., when describing deeply bound Efimov states or three-body 
recombination processes. 

Let us first explicit the connection between our integral equations
in the BCS limit and the 1D reduced problem.
We start with bosons. The 1D Schr\"odinger equation is
\begin{equation}\label{bethe2}
\begin{split}
& \left[ -( \partial^2_{z_1}  + \partial^2_{z_2} + \partial^2_{z_3} ) - 
\frac{2 m_0 E}{\hbar^2} \right] \psi(z_1,z_2,z_3) = \\[2mm]
& \frac{4}{a_{aa}} \left[ \delta(z_1-z_2) + \delta(z_1-z_3) + \delta(z_2-z_3) 
\right]\psi(z_1,z_2,z_3),
\end{split}
\end{equation}
where the two-body 
$1$D contact interaction can be recovered by projecting the $3$D
pseudopotential to the lowest transverse state,
\begin{equation}
-\frac{2 \hbar^2}{m_0 a_{aa}} \delta(z) = \int  \rho d {\bf \rho} d\phi \
R_{00} (\rho) \frac{4\pi\hbar^2 a}{m_0} \delta({\bf r}) 
\frac{\partial}{\partial r} (r \cdot) R_{00} (\rho).
\end{equation}
We then closely follow Sec.~\ref{sec2}, with the
 $1$D boundary condition for $y\to 0$,
\begin{equation}\label{boundary1d} 
\psi (x,y) = f(x) \left( 1- \frac{|y|}{a_{aa}} \right),
\end{equation}
 obtained by projecting Eq.~\eqref{asymto} to the lowest transverse 
state. One finally obtains, 
after some algebra and 
the rescaling $\bar{k} = (\sqrt{3}/2 \, a_{aa}) \bar{u}$, 
the integral equation 
\begin{equation}\begin{split}
& \left[\frac{1}{\sqrt{1+3(u^2-\bar{u}^2)/4}}-1\right]f(u)
\\[2mm]
&=-\frac{2}{\pi}\int_{-\infty}^\infty\frac{du'f(u')}{1+u^2+u'^2 + u u'
-3 \bar{u}^2/4},\label{ieBCS2}\end{split}
\end{equation}
which can also be obtained from the integral equation \eqref{inteqbos}
in the BCS limit, i.e., from the scattering approach.
For fermions, the same line of reasoning can be followed,
and one obtains the integral equation \eqref{ieBCS2} with an overall
factor $-1/2$ factor on the right hand side.
We conclude that the direct solution of Eq.~\eqref{bethe2} can provide
exact results in the BCS limit
 for $a_{ad}$, $b_{ad}$, and possible trimer bound states.

\subsection{Bosonic case}\label{bosonic}

Equation \eqref{bethe2} can be solved by using the Bethe Ansatz.
The corresponding bosonic three-body wavefunction is written
in the fundamental domain $\mathcal{D}_1 = \{ z_1<z_2<z_3\}$ as
\begin{equation}\label{general-bethe}
\psi(z_1,z_2,z_3) = \mathcal{N} 
\sum\limits_P A(P)\,e^{i\sum\limits_{j=1}^3 z_j k_{Pj}},
\end{equation}
where  $\mathcal{N}$ is a normalization constant,
and the sum extends over all
permutations of $\{1,2,3\}$, with coefficients \cite{gaudin}
\begin{equation}\label{coeffA}
A(P)=\prod\limits_{1\le j<l\le3}\left(1+\frac{ 2i/a_{aa}}{k_{Pl}-k_{Pj}}\right).
\end{equation}
In other domains, the wavefunction is recovered by symmetry arguments.
There is only one trimer bound-state solution \cite{castin,straley}, 
\begin{equation}
\psi(z_1,z_2,z_3) = \mathcal{N} \exp \left[ - 2(z_3-z_1)/a_{aa} \right],
\end{equation}
here given in the fundamental domain  $\mathcal{D}_1$,
or in general form by Eq.~\eqref{BAtrimer}. Using the boundary condition
\eqref{boundary1d}, we find
that $f(u) = 1/(u^2+4)$ solves Eq.~\eqref{ieBCS2} with $\bar{u}
= i u_0 = 2 i$, as discussed in Sec.~\ref{trimers}.

We now turn to the atom-dimer scattering problem,
putting $a_{aa}k_1 = \bar{u}$, $a_{aa} k_2 = - i -\bar{u}/2 $ and $a_{aa} k_3 =  i  - \bar{u}/2$, with the
corresponding total energy 
\begin{equation}\label{totener}
E = \frac{\hbar^2}{m_0 \, a_{aa}^2} \, \left[ -1 
+ \frac{3}{4} \, \bar{u}^2 \right],
\end{equation}
see Eq.~\eqref{energy} in the BCS limit.
{}From Eq.~\eqref{general-bethe}, the three-boson wavefunction is
 explicitely given. Using then the $1$D boundary condition
\eqref{boundary1d}, we find in real space
after some algebra, up to an overall normalization constant,
\begin{equation}\label{resultf}
f(x) = (2-3 i \epsilon \bar{u}) 
\left[ - (2 - i \epsilon \bar{u}) e^{i \bar{u} x}
+ 2  (2+ i \epsilon \bar{u}) e^{-i \bar{u} x/2} e^{-|x|} \right],
\end{equation}
where $\epsilon = {\rm sgn} (x)$, and we have performed the
rescaling $\sqrt{3} x / 2 a_{aa} \to x$ in order to give the correct
 atom-dimer distance in units of $a_{aa}$. 
Remarkably, the atom-dimer scattering does not give any reflected wave.
The atom just passes the dimer and only acquires a phase shift, without
being actually backscattered.
Consequently, one can neither use the scattering Ansatz \eqref{skor} nor 
define $a_{ad}$ and $b_{ad}$ as in Eq.~\eqref{aaddef}. We therefore
need a more general definition to take into account the
reflectionless situation  found in the BCS limit.
{}From Eq.~\eqref{resultf} and its
counterpart with $\bar{u} \to -\bar{u}$, we can use a simple but convenient
 basis for the incoming waves. The symmetric (antisymmetric) 
state is taken as a plane wave $e^{i \bar{u} x}$
coming from the left plus a plane wave $e^{-i \bar{u} x}$ ( $-e^{-i \bar{u} x}$)
from the right. In that basis, the
scattering process is described by the $2\times2$
scattering matrix \cite{sutherland}
\begin{equation}\label{Smatrix}
\mathcal{S} \equiv \begin{pmatrix} - e^{i \delta_s (\bar{u})} & 0 \\
0 & e^{i \delta_a (\bar{u})} \end{pmatrix} = 
\begin{pmatrix} t + r & 0 
\\
0 & t-r \end{pmatrix},
\end{equation}
where $r$ and $t$ are again  energy-dependent
 reflection and transmission coefficients that 
can be read off from Eq.~\eqref{resultf}.
At low energy, after expanding in $\bar{k} = \bar{u}/a_{aa}$, 
 the phases $\delta_{a,s}$ define two scattering lengths according to
\begin{equation}\label{aaddef2}
\delta_s(\bar{k}) = - 2 \bar{k}\, a_{ad} + \mathcal{O} (\bar{k}^2), \quad 
\delta_a(\bar{k}) = 2\bar{k}\, b_{ad} + \mathcal{O} (\bar{k}^2).
\end{equation}
This definition is more general than 
Eq.~\eqref{aaddef} and reduces to
it when the Ansatz \eqref{skor} applies.
We also observe that in general two scattering lengths are needed
to completely describe an arbitrary scattering process in 1D. 

Using Eq.~\eqref{resultf}, we find for this atom-dimer scattering 
problem the result $b_{ad} = - 2 a_{aa}=-1/\sqrt{\Omega_B}$. This agrees with the asymptotic
result in Fig.~\ref{aadbad} for the BCS limit.
Furthermore, we find $a_{ad} = -\infty$, in accordance with the
divergence $a_{ad}\propto - 1/\Omega_B^{3/2}$, see Fig.~\ref{aadbad}.
Finally, we infer analytical results
from Eq.~\eqref{resultf} for the integral equation \eqref{inteqfinbos} 
in the BCS limit.
 Switching to Fourier space and expanding $f$
in $\bar{u}$ as $f (u) = f_0(u) + \bar{u} f_1 (u)$, we find
\begin{eqnarray}
\label{eqf0} f_0 &=& -\pi \delta(u) + \frac{2}{1+u^2}, \\
\label{eqf1} f_1 &=& -\pi \delta'(u) + \frac{2}{u (1+u^2)^2},
\end{eqnarray}
where each of these functions satisfies Eq.~\eqref{ieBCS2}
for $\bar{u}=0$.
Plugging $f_1$ into Eq.~\eqref{ieBCS2} for $\bar{u}=0$, we confirm the
asymptotic behavior $b_{ad} =-1/\sqrt{\Omega_B}$ in the BCS limit,
see Eq.~\eqref{bsol}. Plugging $f_0$ into Eq.~\eqref{ieBCS2}
for $\bar{u}=0$, we obtain  the solution \eqref{realsol}
 with $h(0)=0$ for any finite $a_{ad}$. 
Therefore the condition $h(0)=-1$ implies $a_{ad} = -\infty$.

\subsection{Fermionic case}\label{fermionic}

We now consider the fermionic $(\uparrow\uparrow\downarrow)$
 three-body problem and solve it via the Bethe Ansatz. 
The Schr\"odinger equation is essentially Eq.~\eqref{bethe2} with 
the three $\delta$-functions  replaced by 
$\delta(z_1-z_2) + \delta(z_1-z_3)$. This equation does not
lead to any trimer state.
The Bethe Ansatz for fermions is slightly different from the one for 
bosons. The form Eq.~\eqref{general-bethe} also
applies to the fermionic case, but the coefficients $A(P)$ are 
not  given  by Eq.~\eqref{coeffA} anymore. Moreover, the fundamental domain
$\mathcal{D}_1$ is not sufficient, and we need 
to know the wavefunction in one more domain \cite{lieb2}, e.g.,
$\mathcal{D}_2 = \{z_2<z_1<z_3\}$. The wavefunction in other 
domains then follows by  anti-symmetry properties.
 
To study  atom-dimer scattering, we now consider the momenta
$a_{aa}k_1 = - i - \bar{u}/2$,  
$a_{aa}k_2 =  i - \bar{u}/2$, and $a_{aa}k_3 = \bar{u}$,
with total energy \eqref{totener}. Imposing the fermionic anti-symmetry
 and normalizability, the problem reduces 
to the determination of three variables $\alpha_{1,2,3}$.
The wavefunction
is given in $\mathcal{D}_1$ by
\begin{equation}
\psi_1 (z_1,z_2,z_3) = e^{i (k_1 z_1 + k_3 z_2 +  k_2 z_3)}
-  e^{i (k_1 z_1 + k_2 z_2 +  k_3 z_3)},
\end{equation}
and in $\mathcal{D}_2$ by 
\begin{equation}
\begin{split}
\psi_2 (z_1,z_2,z_3) &= \alpha_1 e^{i (k_3 z_1 + k_1 z_2 +  k_2 z_3)}
+ \alpha_2  e^{i (k_1 z_1 + k_3 z_2 +  k_2 z_3)} \\[2mm]
&+ \alpha_3
 e^{i (k_2 z_1 + k_1 z_2 +  k_3 z_3)}.\end{split}
\end{equation}
The variables $\alpha_i$ 
are then obtained by imposing 
boundary conditions for the wavefunction and its first derivative
with respect to $z_1$ and $z_2$ at the boundary between $\mathcal{D}_1$
and  $\mathcal{D}_2$, $\{z_1=z_2 < z_3\}$. The result
is $\alpha_1 = 1- \chi$, $\alpha_2 = \chi$ and $\alpha_3 = -1$,
where 
\[
\chi = -\frac{2+3 i \bar{u}}{2+3 i \bar{u}}.
\]
Once the wavefunction $\psi (z_1,z_2,z_3)$ is known, we can use the
boundary condition \eqref{boundary1d} and the rescaling
$\sqrt{3} x / 2 a_{aa} \to x$ to get
\begin{equation}\label{resultf2}
f(x) = \epsilon (2 - 3 i \epsilon \bar{u} ) \left( e^{-i \bar{u} x/2}
e^{-|x|} - e^{i \bar{u} x} \right),
\end{equation}
where again $\epsilon = {\rm sgn} (x)$.
Similar to the bosonic case, for arbitrary $\bar{u}$,
 there is no reflected wave. The scattering only leads to a phase shift
that goes to $\pi$ at zero energy. This implies a transmission coefficient
$t=-1$, in contrast to the bosonic result $t=1$.
This discrepancy originates from the antisymmetry of the 
fermionic wavefunction with
respect to exchange of the two $\uparrow$ atoms.
As has been discussed in Sec.~\ref{numerical},
we obtain from Eq.~\eqref{resultf2} and the definition \eqref{aaddef2}
that $a_{ad} = (3/2) a_{aa} = 0.75/\sqrt{\Omega_B}$,
 and $b_{ad} = + \infty$. These results confirm the asymptotic behaviors
shown in Fig.~\ref{fig2} for the BCS limit.

Also in the fermionic case, we can obtain analytical results 
for the integral equation \eqref{inteqfin} in  the
BCS limit. We first Fourier transform Eq.~\eqref{resultf2}.
The lowest order in $\bar{u}$ gives 
\begin{equation}\label{eqf02}
f_0 (u) = \frac{i}{u (1+u^2)}.
\end{equation}
Plugged into the fermionic equivalent of \eqref{ieBCS2} for $\bar{u}=0$,
one finds the zero mode responsible for the divergence of $b_{ad}$, see  Sec.~\ref{numerical}. 
More precisely, $h_0 = (-i) u^2 f_0(u)$
satisfies the imaginary part of Eq.~\eqref{inteqfin} in the BCS
limit.
The next order in $\bar{u}$ cannot be taken directly, since the 
resulting integral in the fermionic equivalent of 
Eq.~\eqref{ieBCS2} is not well-defined for $u \to0$
and $\bar{u} \to0$.
Instead, we use that $f(u)$ has the following form
\begin{equation}\label{ansatz2}
f(u) = f_0(u) + \frac{3}{2}\, i \pi \bar{u} \delta(u-\bar{u})
+ i \Gamma (u,\bar{u}) \left( \frac{1}{\bar{u}-u} +\frac{1}{u} \right),
\end{equation}
where $\Gamma (u,\bar{u})$ is a regular function. We insert this
expression into the fermionic equivalent of Eq.~\eqref{ieBCS2}
and look at the $\bar{u} \to0$ limit. After some algebra as in
Sec.~\ref{subsecA}, we obtain 
\begin{equation}\label{res}
\begin{split}
\int_{-\infty}^{\infty}  \frac{d u'}{ 2\pi u'^2} 
[ G(u,u') h(u') -  G(u,0) h(0)]   \\[2mm] 
-  \left( \frac{1}{\sqrt{1+3 u^2/4}}-1\right)
 \frac{h(u) }{2u^2}
 =  \frac{3}{4} \frac{1}{1+u^2} \end{split}
\end{equation}
with 
\begin{equation}\label{resh}
h(u) = \Gamma(u,0) = -1 + \frac{u^2 (2+u^2)}{(1+u^2)^2} = - \frac{1}{(1+u^2)^2}
\end{equation}
and $G(u,u') = (1+u^2+u'^2+u u')^{-1}$.
This shows explicitly that Eq.~(\ref{resh}) is a solution to the real
part of Eq.~\eqref{inteqfin} in the BCS limit, with the expected 
asymptotic behavior $a_{ad}/a_\perp = 0.75/\sqrt{\Omega_B}$. 
Although the form \eqref{ansatz2} seems similar to the general
Ansatz \eqref{skor}, it is different since there is no reflection
in the BCS limit, and Eq.~\eqref{skor} cannot apply.

\subsection{Reflectionless potential}\label{secrefk}

We have seen above that the divergence of $b_{ad}$ ($a_{ad}$) for
fermions (bosons) in the BCS limit was due to the reflectionless character of 
the atom-dimer effective potential. We shall now generalize this
idea and show that for 1D scattering processes of a particle
by any (possibly non-local) symmetric potential, the following
statements are equivalent: (i) there is a divergence of the
1D scattering length $a_1$ 
(the `odd' scattering length $b_1$), (ii) the potential
has a quasi-bound state with even (odd) parity 
precisely at zero energy, and (iii) there is no reflection at zero 
energy. The atom-dimer scattering problem in the BCS limit, 
where higher channels are negligible, is in fact equivalent to a 1D scattering
problem with such an effective non-local potential \cite{prl}.
Therefore, the considerations in this section directly
lead to physical insights for the BCS limit.
The fact that in our case the reflection coefficient is 
zero for {\sl any} incoming energy is an additional feature that cannot
be inferred from the behavior of $a_1$ and/or $b_1$.

Symmetric potentials imply the low-energy 
scattering properties \cite{landau}
\begin{equation}\label{scalow}
t(\bar{k})|_{\bar{k}\to0}=-\frac{2i\bar{k}}{\nu},\quad r(0)=-1,
\end{equation}
where $\bar{k}$ is the wavevector of the incoming plane wave, and $\nu\ne 0$
is a {\sl real}\ parameter that depends on the potential. 
The incoming wave is then totally reflected at zero energy.
There is only one exception to this general low-energy behavior, arising
for potentials with $\nu=0$.
In that case, the limit $\bar{k} \to 0$ is different and characterized by 
\begin{equation}\label{norefle}
t(0)= t_o = \pm 1,\quad r(0)= r_o = 0,
\end{equation}
corresponding to a reflectionless situation at zero energy.
For very small but finite $\nu$, Eq.~\eqref{scalow}
still holds, but the reflection and transmission amplitudes change
on the scale of unity in a narrow momentum region $\bar{k} \sim \nu$.
For $\bar{k}>\nu$ (but still smaller 
than all other momentum scales), Eq.~\eqref{norefle} is reached,
and the potential is effectively reflectionless.
We now show that for $\nu=0$, the conditions (i), (ii) and (iii)
are satisfied and equivalent to each other,
 while for finite $\nu$, strictly speaking, none of them holds.

We first observe from the general definitions of $a_1$ and $b_1$, 
see Eqs.~\eqref{Smatrix} and \eqref{aaddef2}, 
that Eq.~\eqref{norefle} leads to the divergence of $a_1$ ($b_1$) 
for $t_o=1$ ($t_o=-1$). For $\nu\ne0$, $a_1$ and $b_1$ are finite so
that (i) holds iff (iii) is true.
The small-$\nu$ regime is also interesting to investigate.
Using Eq.~\eqref{scalow}, we find that for $t(\bar{k}\sim\nu) = 1$~($ -1$), 
$a_1$ ($b_1$) diverges as $-2/\nu$. To see if there is a quasi-bound
state at zero energy, we consider the asymptotic form of the
wavefunction
\begin{equation}\label{phias}
\psi(x)=\left\{
\begin{array}{ll} e^{i\bar{k}x}+r(\bar{k})
e^{-i\bar{k}x}, & x\to-\infty\;, \\
t(\bar{k})\, e^{i\bar{k}x}, & x\to+\infty\;,
\end{array}\right.
\end{equation}
for which we take $\bar{k}=0$. If $\nu\ne0$, Eq.~\eqref{scalow} gives
$\psi = 0$ in the asymptotic region. Since a quasi-bound state at zero
energy can only be an extended state, we conclude that $\psi$ vanishes 
identically in this case. Conversely for $\nu=0$, 
we obtain $\psi(x) = 1$ for $x\to-\infty$ and $\psi(x) = t_o = \pm1$ 
for $x\to+\infty$. This means that there is an extended quasi-bound 
state at zero energy. Its parity follows from the asymptotic behavior,
and we find that it is even if $t_o=+1$ but odd for $t_o=-1$.
This shows that (ii) and (iii), and hence also (i), are equivalent.
 
In our specific case, the zero energy mode is found by taking 
$\bar{u}\to0$ in Eqs.~\eqref{resultf} and \eqref{resultf2}. This 
leads to (again, $\epsilon={\rm sgn}(x)$)
\begin{equation}
f_0(x) = \left\{ \begin{array}{ll} 4 \left( 2 e^{-|x|} -1 \right),
 & \textrm{for bosons}, \\
2 \epsilon \left( e^{-|x|}-1 \right), & \textrm{for fermions},
\end{array}\right.
\end{equation}
or Eqs.~\eqref{eqf0} and \eqref{eqf02} in momentum space. Now 
$f_0(x)$ is
an even (odd) function for bosons (fermions), corresponding to a divergence
of $a_1$ ($b_1$), as expected from our discussion above. 

To conclude, let us discuss the position of the
quasi-bound state for large $a_1$ or $b_1$ 
by analyzing the poles of the scattering
matrix $\cal{S}$ in Eq.~\eqref{Smatrix}. If a pole appears for the
scattering amplitude $e^{i \delta_s (\bar{k})}$ ($e^{i \delta_a (\bar{k})}$),
the corresponding bound state is even (odd). For large values of $a_1$,
we can use the expansion \eqref{aaddef2}, and the equation for
the even bound state is given by
\begin{equation}
1 + i \bar{k} a_1 =0.
\end{equation}
Analytic continuation to the physical sheet $\bar{k} = i \kappa$ 
($\kappa>0$) gives a real bound state with $\kappa = 1/a_1$ for $a_1>0$,
with energy $E_e = -\hbar^2/m_0 a_1^2$. 
However, for $a_1<0$, we find,
by analytic continuation to the unphysical sheet $\bar{k} = -i \kappa$,
a {\sl virtual}\ bound state with positive energy $E_e = \hbar^2/m_0 a_1^2$. 
For the three-boson problem in the BCS limit, $a_{ad} \to -\infty$. 
Consequently, the virtual quasi-bound state here comes down to zero
from positive energies,
and thus never becomes a real bound state of the three bosons.
 It simply reaches
zero energy in the BCS limit, leading to the divergence of $a_1$.
The situation is similar for an odd bound state. With the expansion
 Eq.~\eqref{aaddef2}, we find the bound-state equation
$1 - i \bar{k} b_1 =0$, so that the same conclusions as above can be 
formulated, with just a sign difference for $b_1$. 
Concerning the three-fermion
problem, the same scenario for the virtual bound state is encountered,
since there $b_{ad} \to+\infty$ in the BCS limit.

\section{Conclusions}
\label{conc}

 In this paper, we presented the results of our study of the 
three-body problem in a quasi-1D confinement.
We define and calculate two different 1D atom-dimer scattering lengths,
 $a_{ad}$ and $b_{ad}$, which are directly accessible in scattering
experiments. Physically, $a_{ad}$ reflects the low-energy properties
 of the scattering phase shift for a symmetric wave while  $b_{ad}$
is the equivalent for an antisymmetric wave.
 Both can be inferred from the energy dependence
of the transmission and the reflection coefficients.
 Technically, we derive a system of integral equations for the 1D 
scattering amplitudes with different transverse channel indices. 
For the sake of simplicity, the above system is projected onto
 the transverse ground state. We solve the ground-state equation 
 numerically in general and analytically in the dimer and BCS limits.  
Note that the dimer and the BCS limits each correspond to different
approaches to the 3D problem ($|a|\ll a_\perp$). Therefore we have
investigated in detail the role of the higher transverse channels 
in both cases. 

In the dimer limit, indeed it turns out that the higher channels contribute
on the same footing. Their role is described by the 3D equations
derived   previously \cite{skorniakov}.
Note that the definition of the 1D scattering length
remains distinct from that of the 3D scattering length even in the 
dimer limit. We establish a simple analytic relation between the latter
and $a_{ad}$, valid both for bosons and fermions. In this limit $b_{ad}\to0$,
indicating that atom-dimer scattering can be regarded as potential
scattering with a short-range effective potential. The resulting 
$a_{ad}$ is negative for fermions but positive for bosons pointing
 to the existence of the confinement-induced trimer state in the 
latter case. Indeed we
find such a state from the bound-state equation for bosons. Its energy
is independent of the bosonic regularization parameter $R^*$ in the
large $R^*$ limit.

The BCS limit is even more interesting: we show that higher channels can
always be neglected here as far as the low-energy scattering is concerned.
We find that the scattering is described by the three-particle Bethe Ansatz
equations, i.e., we are dealing with two-body contact 
interactions. Then atom-dimer scattering cannot be viewed as
potential scattering anymore. Although an effective potential can be defined,
it is non-local and  not short-ranged \cite{prl}. Instead atom-dimer
scattering is found to be reflectionless: $a_{ad}$ diverges for bosons,
while $b_{ad}$ diverges for fermions. In fact the Bethe Ansatz
equations remain applicable also for the $N$-body problem in the
BCS limit. 

We have found a novel confined-induced trimer state for bosons.
We trace the trimer state numerically from the dimer to the 
BCS limit provided Efimov physics can be neglected. Specifically 
in the dimer limit, we have discussed the interplay 
between this confined trimer state and the usual Efimov bound states.
 The trimer state is unique, its energy is nearly universal,
and it matches the known Bethe Ansatz three-particle bound state in
the BCS limit.

\acknowledgments
We thank A. Komnik for discussions.
This work was supported by the SFB TR12 of the DFG.

\appendix
\section{}\label{appen}

We show here how to obtain the integral representation Eq.~\eqref{zeta} from the more
standard  expression (see Ref.~\cite{moore2} for example)
\begin{equation}
\zeta(1/2,\Omega) = \lim_{N\to+\infty} \left( \sum_{n=0}^{N} \frac{1}{(n+\Omega)^{1/2}}    - 2 (N+\Omega)^{1/2} 
\right).
\end{equation}
We define $A_N = \left( \sum_{n=0}^N ( n+ \Omega)^{-1/2} \right) -2 \sqrt{N+\Omega}$ such that 
$\lim_{N\to +\infty} A_N = \zeta(1/2,\Omega)$. Using the integral representations
\[
\begin{split}
\frac{1}{\sqrt{n+\Omega}} &= \int_{0}^{\infty} 
\frac{dt}{\sqrt{\pi t}} e^{-(n+\Omega) t}, \\[2mm]
 \sqrt{N+\Omega} &= 
\int_{0}^{\infty} \frac{dt}{\sqrt{\pi t}} \frac{1 - 
e^{-(N+\Omega) t}}{t},
\end{split}
\]
and the geometrical summation $\sum_{n=0}^N e^{-n t} = (1 - e^{ -(N+1) t})/(1-e^{-t})$,
$A_N$ can be written as
\begin{equation}\label{a2}
\begin{split}
A_N = \int_0^\infty & \frac{dt}{\sqrt{\pi t}}
\left( \frac{e^{-\Omega t}}{1-e^{-t}} - \frac{1}{t} \right) \\[2mm]
&+ \int_0^\infty \frac{dt}{\sqrt{\pi t}}
\left( \frac{1}{t} - \frac{e^{-t}}{1-e^{-t}} \right)
e^{-(N+\Omega) t}.
\end{split}
\end{equation}
The second term in Eq.~\eqref{a2} vanishes for large $N$ so that we obtain the integral
representation of Eq.~\eqref{zeta}.

\end{document}